\documentclass[preprints,article,accept,pdftex,oneauthor]{Definitions/mdpi} 
\def\simge{\mathrel{\rlap{\raise 0.511ex
     \hbox{$>$}}{\lower 0.511ex \hbox{$\sim$}}}}
\def\simle{\mathrel{\rlap{\raise 0.511ex
      \hbox{$<$}}{\lower 0.511ex \hbox{$\sim$}}}}
\firstpage{1} 
\pubvolume{1}
\issuenum{1}
\articlenumber{0}
\pubyear{2023}
\copyrightyear{2022}
\externaleditor{Academic Editor: Armen Sedrakian}
\datereceived{1 November 2022} 
\dateaccepted{26 December 2022} 
\datepublished{4 January 2023} 
\hreflink{https://doi.org/} % If needed use \linebreak
\Title{Constraints on Nuclear Symmetry Energy Parameters}

\TitleCitation{Constraints on Nuclear Symmetry Energy Parameters}

\Author{James M. Lattimer \orcidA{}%, Firstname Lastname $^{2,\ddagger}$ and Firstname Lastname $^{2,}$*
}

\AuthorNames{James M. Lattimer}

\AuthorCitation{Lattimer, J. M.}
\address[1]{%
Department of Physics \& Astronomy, Stony Brook University, Stony Brook, NY 11794-3800, USA; james.lattimer@stonybrook.edu}

\abstract{A review is made of constraints on the nuclear symmetry energy parameters arising from nuclear binding energy measurements, theoretical chiral effective field predictions of neutron matter properties, the unitary gas conjecture, and measurements of neutron skin thicknesses and dipole polarizabilities. While most studies have been confined to the parameters $S_V$ and $L$, the important roles played by, and constraints on $K_{\rm sym}$, or, equivalently, the neutron matter incompressibility $K_N$, are discussed. Strong correlations among $S_V, L$, and $K_{N}$ are found from both nuclear binding energies and neutron matter theory.  However, these correlations somewhat differ in the two cases, and those from neutron matter theory have smaller uncertainties. To 68\%
confidence, it is found from neutron matter theory that $S_V=32.0\pm1.1$ MeV, $L=51.9\pm7.9$ MeV and $K_N=152.2\pm38.1$ MeV.  Theoretical predictions for neutron skin thickness and dipole polarizability  measurements of the neutron-rich nuclei $^{48}$Ca, $^{120}$Sn, and $^{208}$Pb are compared to recent experimental measurements, most notably the CREX and PREX neutron skin experiments from Jefferson Laboratory. By themselves, PREX I+II measurements of $^{208}$Pb and CREX measurement of $^{48}$Ca suggest $L=121\pm47$ MeV and $L=-5\pm40$ MeV, respectively, to 68\% confidence. However, we show that nuclear interactions optimally satisfying both measurements imply $L=53\pm13$ MeV, nearly the range suggested by either nuclear mass measurements or neutron matter theory, and is also consistent with nuclear dipole polarizability measurements. This small parameter range implies $R_{1.4} = 11.6\pm1.0$ km and $\Lambda_{1.4} = 228^{+148}_{-90}$, which are consistent with NICER X-ray and LIGO/Virgo gravitational wave observations of neutron stars.}


\begin{document}

%%%%%%%%%%%%%%%%%%%%%%%%%%%%%%%%%%%%%%%%%%
%\endnote{This is an endnote.} % To use endnotes, please un-comment \printendnotes below (before References). Only journal Laws uses \footnote.

% The order of the section titles is: Introduction, Materials and Methods, Results, Discussion, Conclusions for these journals: aerospace,algorithms,antibodies,antioxidants,atmosphere,axioms,biomedicines,carbon,crystals,designs,diagnostics,environments,fermentation,fluids,forests,fractalfract,informatics,information,inventions,jfmk,jrfm,lubricants,neonatalscreening,neuroglia,particles,pharmaceutics,polymers,processes,technologies,viruses,vision

\section{Introduction}

The nuclear symmetry energy, and 
its density dependence, as characterized by the traditional symmetry energy parameters $S_V$ and $L$, has been the focus of much recent activity because it is the most direct link between nuclear physics and nuclear astrophysics~\cite{Steiner2005,LP07,Lattimer2021}.  Both the expected neutrino and gravitational wave
signals from gravitational collapse supernovae within our Galaxy are
sensitive to the symmetry energy~\cite{SLM94, Roberts12, Morozova2018}.  The
symmetry energy near  the baryon density at saturation, $n_s=0.155\pm0.005$ fm$^{-3}$, determines
the radius~\cite{LP01} of a neutron star (NS), which strongly
influence the expected gravitational signals from their mergers~\cite{Bauswein12,Lackey2012}.  The symmetry energy also affects
the NS crust's thickness and thermal relaxation time, potentially observable in
cooling and accreting~\cite{PR12} NSs and in giant magnetar
flares~\cite{Hurley2005,Thompson2001,Samuelsson2007}. The composition of matter at densities above
$n_s$, and the existence of neutrino processes which can rapidly cool
NSs, depend on the density dependence of the symmetry energy~\cite{LPPH91}, as do
predicted properties of neutron-rich nuclei and reaction rates
involved in the astrophysical r-process~\cite{Nikolov11}. 

Experimental attempts to constrain the nuclear
symmetry energy parameters include measurements of nuclear masses, neutron skin thicknesses,
nuclear dipole polarizabilities, giant and pygmy dipole resonance energies,
flows in heavy-ion collisions, and isobaric analog states.  These
constraints are influenced by varying degrees of model dependence. 
In addition, recent advances in neutron matter theory, especially that of systematic expansions involving chiral effective field theory ($\chi$EFT)~\cite{Drischler2021}, also constrain the symmetry energy parameters. 

Of considerable interest are the recent parity-violating electron scattering neutron skin experiments of $^{208}$Pb  (PREX-I and PREX-II) ~\cite{Adhikari2021} and $^{48}$Ca~\cite{Adhikari2022}.  These measure the mean square difference of the neutron and proton radii using a technique which is argued to be the most direct and least model-dependent experiment to date~\cite{Thiel2019}. PREX I+II combined yields $r_{np}^{208}=0.283\pm0.071$ fm~\cite{Adhikari2021},  which implies ~\cite{Reed2021} 68\% confidence ranges of $S_V=38.29\pm4.66$ MeV and $L=109.56\pm36.41$ MeV. Both values, and the measured value of $r_{np}^{208}$ itself, are considerably larger than from expectations from neutron matter and nuclear binding energies, and also from previous measurements, although overlapping with them at about the 90\% confidence level.  This indicates a tension with the current understanding of the equation of state (EOS).  For example, these results imply a tidal deformability that lies above the 90\% confidence upper limit established for a $1.4M_\odot$ NS by the LIGO/Virgo observation of the binary NS (BNS) merger GW170817 ~\cite{De2018,Abbott2019}.   In contrast, the measurement of the neutron skin of $^{48}$Ca using the same technique~\cite{Adhikari2022} are somewhat smaller than the average of earlier experimental measurements and expectations from nuclear binding energies and neutron matter theory.

Ref.~\cite{Zhang2022} performed a Bayesian analysis of the PREX and CREX results.  They found that the two experimental results are incompatible with each other at 68\% confidence level, but compatible at 90\% confidence level.  Combining the data, they inferred $S_V=30.2^{+4.1}_{-3.0}$ MeV and $L=15.3^{+46.8}_{-41.5}$ MeV at 90\% confidence level.  They find the combined results predict $r_{np}^{48}$ close to the CREX result, but predict $r_{np}^{208}$ considerably smaller than the PREX result.  Ref.~\cite{Reinhard2022} also performed a combined analysis, and conclude that a simultaneous accurate description of the skins of $^{48}$Ca and $^{208}$Pb cannot be achieved with their models that accommodate mass, charge radii and experimental dipole polarizabilities.

In this paper we take a different perspective by discovering the properties of nuclear interactions fit to binding energy and charge radii of large numbers of nuclei that best satisfy both PREX and CREX measurements. We agree with the assessments of both Ref.~\cite{Zhang2022,Reinhard2022} that no conventional nuclear interaction can fit both measurements to 68\% confidence level.  However, our optimum fit results in ranges of $S_V$ and $L$ that not only have central values larger than those estimated by Ref.~\cite{Zhang2022} but also smaller uncertainties.  Our results also agree with results from neutron matter theory while those from Ref.~\cite{Zhang2022} do not.  We find  similar results when historical measurements of the neutron skins of both nuclei are utilized, instead, suggesting that systematic uncertainties in the measurements are not dominant. 

We begin by summarizing nuclear binding energy and theoretical neutron matter constraints on the nuclear symmetry energy parameters $S_V, L$ and $K_N$.  We show that estimates of symmetry energy parameters from chiral Lagrangian expansions of nuclear matter are more reliably estimated from neutron matter calculations than from both neutron and symmetric matter calculations.  We explore systematic uncertainties in parameter estimation stemming from the choice of nuclear interaction model, and note apparent inconsistencies associated with relatively stiff relativistic mean field (RMF) interactions.  We show that nuclear models fit to nuclear binding energies that optimally satisfy both CREX and PREX neutron skin measurements confine symmetry parameter values to narrow ranges that are consistent with expectations from neutron matter theory.  We also show that they are also consistent with theoretical estimates based on dipole polarizability experiments.  We also compare our estimates of symmetry parameters from those estimated from astrophysical observations of neutron stars, especially from gravitational wave observations of GW170817 and Neutron Star Interior Composition ExploreR (NICER) X-ray observations of PSR J0030+0451~\cite{Miller2019,Riley2019} and PSR J0740+6620~\cite{Miller2021,Riley2021}.

 %%%%%%%%%%%%%%%%%%%%%%%%%%%%%%%%%%%%%%%%%%
\section{The Nuclear Symmetry Energy}
The nuclear symmetry
energy $S(n)$ is defined here to be the difference between the energies of pure neutron matter (PNM), $E_{N}$, and 
isospin symmetric nuclear matter (SNM), $E_{1/2}$, at the baryon density $n$.  
Related quantities are the density-dependent coefficients, $S_n(n)$, of an expansion of the bulk energy per baryon, $E(n,x)$, in powers of the neutron excess $1-2x$:
\begin{equation}
E(n,x)=E_{1/2}(n)+S_2(n)(1-2x)^2+S_3(n)(1-2x)^3 + \dots
\label{eq:symm}
\end{equation}
A common approximation is to retain only the quadratic term in
Equation~(\ref{eq:symm}) at every density, even for small proton fractions $x$, so that
$S(n)\simeq S_2(n)$.  Chiral Lagrangian expansions for PNM, nuclear matter with admixtures of protons, and SNM, indicate that this approximation appears
valid~\cite{Wellenhofer2016} for all values of $x$.  For matter with densities below $n_s$, such as that in nuclei, experimental evidence for higher-than-quadratic contributions is lacking, but this could be partly due to the near-symmetric character of nuclei. 
It is customary to introduce the volume symmetry energy
 $S_V=S_2(n_s)$, symmetry slope $L=3n_s(dS_2/dn)_{n_s}$, symmetry incompressibility
$K_{sym}=9n_s^2(d^2S_2/dn^2)_{n_s}$, and symmetry skewness $Q_{sym}=27n_s^3(d^3S_2/dn^3)_{n_s}$  parameters, the coefficients of a Taylor expansion in density around $n_s$:
\begin{equation}
S_2=S_V+{L\over3}(u-1)+{K_{sym}\over18}(u-1)^2+{Q_{sym}\over162}(u-1)^3+\cdots,
\label{eq:s}\end{equation}
where $u=n/n_s$.
  If only the quadratic term in Equation~(\ref{eq:symm}) is retained, we note that $S=S_2$.   As a result, the energy per baryon $E_N$ and the pressure $P_N$ of PNM at $n_s$ become
  \begin{equation}
  E_N(n_s)=E(n_s,0)=S_V-B;\qquad P_N(n_s)=P(n_s,0)=Ln_s/3,
  \label{eq:s1}\end{equation}
  where $B\equiv-E_{1/2}(n_s)=16\pm1$ MeV is the bulk binding energy parameter of SNM.

We also introduce the incompressibility and skewness parameters for SNM, $K_{1/2}$ and $Q_{1/2}$, and for PNM, $K_N$ and $Q_N$, respectively, so that
\begin{eqnarray}
E_{1/2}=E(u,1/2)&=&-B+{K_{1/2}\over18}(u-1)^2+{Q_{1/2}\over162}(u-1)^3+\cdots,\cr
E_N=E(u,0)&=&{L\over3}(u-1)+{K_N\over18}(u-1)^2+{Q_N\over162}(u-1)^3+\cdots.
\label{eq:s2}\end{eqnarray}
The corresponding parameters for the symmetry energy are $K_{sym}=K_N-K_{1/2}$ and $Q_{sym}=Q_N-Q_{1/2}$.  $K_{1/2}\simeq230\pm20$ MeV has been deduced from giant monopole resonances~\cite{Agrawal2003,Todd-Rutel2005}, but there is little direct experimental evidence for $Q_{1/2}, K_N$ or $Q_N$. 

In this section, we explore correlations involving the symmetry parameters that arise, experimentally, from fitting nuclear binding energies, and, theoretically, from recent neutron matter theory predictions.  Neither of these methods can alone predict values of $S_V$ or $L$ to high precision.  However, since these correlations are motivated by different considerations, combining them can yield additional restrictions.

\subsection{Nuclear Mass Fitting\label{sec:nuc-mass}}
It is straightforward to understand why a strong correlation between $S_V$ and $L$ from fitting nuclear masses exists by using the simple nuclear liquid drop model, which consists of five main terms,
\begin{equation}
    E_{LD}(A,I)=[-B+S_VI^2]A+[E_S-S_SI^2]A^{2/3}+E_{Coul}.
    \end{equation}
    Here $I=1-2Z/A$ and the individual terms represent the volume, surface, and Coulomb energies, respectively.  Additionally, one should consider shell and pairing energies.
The terms proportional to $I^2$ represent the symmetry energy of a nucleus:
\begin{equation}
    {\cal S}_{LD}(A,I)\simeq S_VAI^2(1-S_sA^{-1/3}/S_V).
    \end{equation}
If the Coulomb energy is ignored, the experimental symmetry energy ${\cal S}_{exp}(A,I)$ can be found by taking half the difference of the measured energies $E_{exp}(A,I)$ of nuclei having the same mass but values of $Z$ and $N$ each differing by 2 units.  This procedure also effectively eliminates shell and pairing effects.

 The optimum values of the parameters $S_V$ and $S_S$ can be found by minimizing
    \begin{equation}
        \chi^2=\sum_i^{\cal N}{[{\cal S}_{exp}(A_i,I_i)-{\cal S}_{LD}(A_i,I_i)]^2\over{\cal N}\sigma^2}
        \end{equation}
with respect to themselves.  ${\cal N}$ is the number of measured nuclei, and $\sigma\sim1$ MeV is a fiducial uncertainty.  The result is a confidence ellipse centered at $S_{V}$ and $S_{S}$ with uncertainties, angle with respect to the $S_S$ axis, and correlation coefficient
\begin{equation}
    \sigma_V=\sqrt{\chi^{-1}_{VV}},\quad \sigma_S=\sqrt{\chi^{-1}_{SS}},\quad \alpha={1\over2}\tan^{-1}{2\chi_{VS}\over\chi_{SS}-\chi_{VV}},\quad  r={\chi_{VS}\over\sqrt{\chi_{VV}\chi_{SS}}},
\end{equation}
respectively, where $\chi^{-1}$ is the matrix inverse of  $\chi_{ij}=\partial^2\chi^2/\partial S_i\partial S_j$.  The slope of the confidence ellipse is $dS_S/dS_V=\cot\alpha\simeq\sigma_S/\sigma_V$ when $\alpha$ is small.  Since ${\cal S}_{LD}$ is linear in $S_V$ and $S_S$, the symmetric matrix $\chi^2$ depends only on the measured $A_i$ and $I_i$, not on $S_{V}$ or $S_{s}$,
\begin{equation}
    \left[\chi_{VV},\chi_{SV},\chi_{SS}\right]={2\over{\cal N}\sigma^2}\sum_iI_i^4\left[A_i^2,-A_i^{5/3},A_i^{4/3}\right]\simeq{1\over\sigma^2}[61.6,-10.7,1.87].
\end{equation}
Numerical values were obtained by using the set of 2336 nuclei from Ref.~\cite{Audi2003} with $N\ge40$ or $Z\ge40$.  As a result, one finds $\sigma_V=2.3\sigma$, $\sigma_S=13.2\sigma$, $\alpha\simeq9.8^\circ$ and $r=0.997$, which represents a high degree of correlation.  

To convert this correlation into one involving $S_V$ and $L$, it can be noted that $S_S$ originates from an integration of the density-dependent symmetry energy through the nuclear density profile.  In the plane-parallel approximation, it can be shown~\cite{Steiner2005} that
\begin{equation}
    S_S={E_SS_V\over2}{\int_0^1\sqrt{u}(S_V/S(u)-1)(E(u,1/2)+B)^{-1/2}du\over\int_0^1\sqrt{u}(E(u,1/2)+B)^{1/2}du}
\end{equation}
The simple approximations $S(u)\simeq S_V+L(u-1)/3$ and $E(u,1/2)\simeq-B+K_{1/2}(u-1)^2$ lead to
\begin{equation}
    {S_S\over S_V}\simeq{135E_S\over 2K_{1/2}}\left[1-\left[{1\over a}-1\right]^{1/2}\tan^{-1}\left(\left[{1\over a}-1\right]^{-1/2}\right)\right].
\end{equation}
where $a=L/(3S_V)$.  When $a\simeq2/3$ and $135E_S/(2K_{1/2})\simeq5$, one finds that $S_S/S_V\simeq1.62$ and $d(S_S/S_v)/da\simeq3.85$.  As a result, $S_S$ increases rapidly with $L$, and the steep $S_S-S_V$ correlation translates into a steep $L-S_V$ correlation, $dL/dS_v\simeq6$, with a similar correlation coefficient.  The liquid droplet model~\cite{Myers1969}, in which the nuclear symmetry energy enters as
\begin{equation}
    S_{LD}(A,I)=AI^2S_V(1+S_SA^{-1/3}/S_V)^{-1},
\end{equation}
provides a much improved
fit, and also shows a  significant correlation between $S_V$ and $L$.

\begin{figure}[H]
\vspace*{-0.5cm}
\hspace*{-1.5cm}\includegraphics[width=15 cm,angle=180]{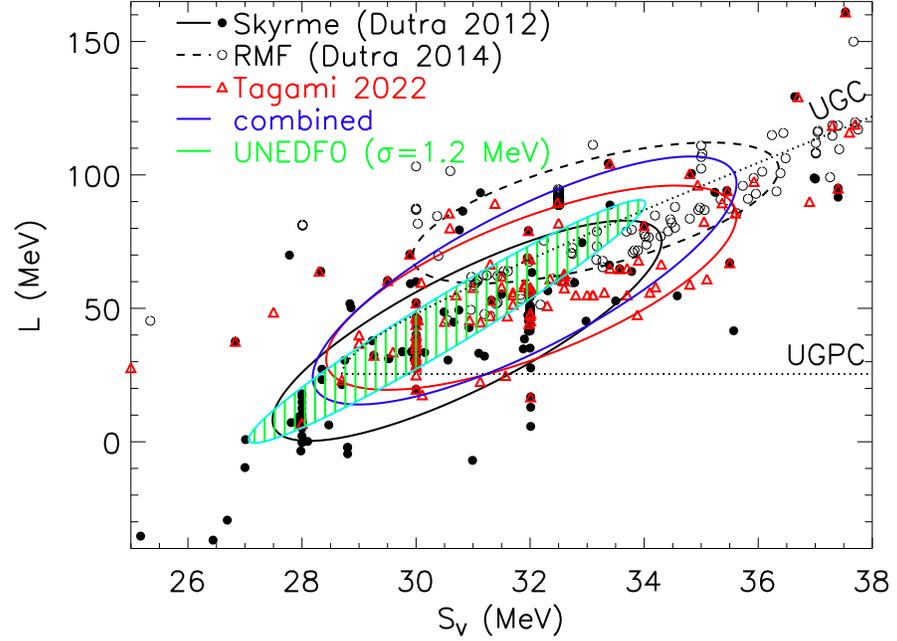}
\vspace*{-0.5cm}
\caption{$S_V$ and $L$ data from individual Skyrme (black filled circles, Ref.~\cite{Dutra2012}), relativistic mean field (RMF, black open circles, Ref.~\cite{Dutra2014}) forces, both interaction types (Tagami 2022, red triangles, Ref.~\cite{Tagami2022}), and all tabulated interactions (combined); corresponding 68.3\% confidence ellipses are shown.  The green hatched confidence ellipse is taken from the UNEDF collaboration~\cite{Kortelainen2010} using $\sigma=1.2$ MeV (see text).  The bounds provided by the Unitary Gas Conjecture (UGC, \cite{Tews2017}) and the Unitary Gas Pressure Conjecture are shown as dotted curves.}
\label{fig:correl0}
\end{figure}

This correlation naturally appears when comparing large numbers of non-relativistic Skyrme-like and RMF nuclear interactions which were fitted to nuclear binding energies and, in some cases, additional properties, such as charge radii.  Of the 240 Skyrme-like interactions studied by Ref.~\cite{Dutra2012}, 45 can be rejected~\cite{Tews2017} since they have some saturation properties (including $K_{1/2}$) outside the empirical window or other anomalous behavior.  Similarly, of the 256 RMF interactions studied by Ref.~\cite{Dutra2014}, 100 can be rejected~\cite{Tews2017}. The compilation of
Ref.~\cite{Tagami2022} contains an additional 206 interactions of both
types, of which 169 survive the conditions imposed by
Ref.~\cite{Tews2017}, but an additional 58 of these are duplicates
from Refs.~\cite{Dutra2012,Dutra2014}. The properties of confidence ellipses for these three groups of interactions are shown in Table~\ref{tab:svldata}.

The UNEDF collaboration~\cite{Kortelainen2010} determined an $S_V-L$ correlation in a more precise fashion  using a universal energy density functional fit to binding energies and charge radii of selected closed-shell nuclei.  The confidence ellipse size depends on the arbitrary value of a fiducial uncertainty parameter $\sigma$.   The value $\sigma\simeq1.2$ MeV yields an approximately equal uncertainty to that of the Skyrme forces in the compilation of Ref.~\cite{Dutra2012}, as seen in Figure~\ref{fig:correl0}.
This is not surprising given the fact that the universal energy density functional of Ref.~\cite{Kortelainen2010} is non-relativistic.  However, the correlation is much tighter than those found from the compilation of Refs.~\cite{Dutra2012,Dutra2014,Tagami2022} possibly because the latter forces were not subjected to the same strict calibration involving charge radii.  Note that the slope and best-fit parameter values $S_{V0}$ and $L_0$ do not depend on the parameter $\sigma$.  Table~\ref{tab:svldata} gives the confidence ellipse specifics.

\begin{table}[H]
\caption{Symmetry energy parameter correlations.\label{tab:svldata}}
%	\begin{adjustwidth}{-\extralength}{0cm}
%		\newcolumntype{C}{>{\centering\arraybackslash}X}
		\begin{tabularx}{\textwidth}{cCCCCC}
\toprule
\textbf{Method/\boldmath$S_V-L$}	& \textbf{\boldmath$S_{V0}$ (MeV)}& \textbf{\boldmath$L_0$ (MeV)} & \textbf{\boldmath$\sigma_{S_V}$ (MeV)} & \textbf{\boldmath$\sigma_L$ (MeV)}& \boldmath$r$\\
		\midrule			UNEDF~\cite{Kortelainen2010}		& 30.5			& 45.1			& 1.9  & 24.0 & 0.970\\
			 Skyrme~\cite{Dutra2012}		& 30.9		& 	41.5		& 2.25 & 27.2 & 0.812\\
			RMF~\cite{Dutra2014}		& 33.1	& 85.8			& 2.12 & 17.4 & 0.625\\
	Tagami~\cite{Tagami2022}		& 32.0		& 57.7			& 2.37 & 25.2 & 0.702\\
Combined~\cite{Dutra2012,Dutra2014,Tagami2022}		& 32.1		& 62.2			& 2.45 & 30.6 & 0.783\\
Combined + UGC/UGPC		& 32.5		& 57.7			& 2.09 & 20.7 & 0.920\\
			$\chi$EFT (SNM+PNM) ~\cite{Drischler2020}		& 31.7			&	59.8		& 1.1 & 4.2 & 0.715 \\
			$\chi$EFT (SNM+PNM)		& 	31.7		& 60.4			& 2.4  & 8.1 & 0.913\\
			$\chi$EFT  (PNM)		& 32.0			& 51.9			& 1.1  & 7.9 & 0.978\\
			neutron skin (CREX+PREX) & 32.2 & 52.9 & 1.7 & 13.2 & 0.820\\
			neutron skin (other) & 31.0 & 42.1 & 1.2 & 8.2 & 0.729\\ 

				\midrule

	\textbf{Method/\boldmath$K_N-L$}	& \textbf{\boldmath$K_{N0}$ (MeV)}& \textbf{\boldmath$L_0$ (MeV)}     & \textbf{\boldmath$\sigma_{K_N}$ (MeV)}& \textbf{\boldmath$\sigma_L$ (MeV)}& \boldmath$r$\\[1pt]
			\midrule
			Skyrme~\cite{Dutra2012}		& 73.3		& 	41.6	&	98.9 & 27.2  & 0.952\\
			RMF~\cite{Dutra2014}		& 234.0 	& 	85.8		& 63.6 & 17.4& 0.666\\
	Tagami 2022~\cite{Tagami2022}& 161.9 & 57.9 & 99.5 & 25.6 & 0.757\\
Combined~\cite{Dutra2012,Dutra2014,Tagami2022}		& 147.7			& 60.4			& 113.7 & 30.6 & 0.899\\
Combined~\cite{Dutra2012,Dutra2014,Tagami2022} +UGC/UGPC		& 137.3			& 57.7			& 74.8 & 20.7 & 0.745\\
			$\chi$EFT (SNM+PNM)		& 172.1			& 60.4			& 27.4 	& 8.1			& 0.558\\
			$\chi$EFT  (PNM)		& 152.3			& 51.9	& 35.1	& 7.9			& 0.993 \\
			neutron skin (CREX+PREX) & 141.6 & 52.9 & 73.2 & 13.2 & 0.530\\
			neutron skin (other) & 104.8 & 42.1 & 75.4 & 8.2 & 0.590\\ 
   \midrule
\textbf{Method/\boldmath$Q_N-L$}	& \textbf{\boldmath$Q_{N0}$ (MeV)}	& \textbf{\boldmath$L_0$ (MeV)}    & \textbf{\boldmath$\sigma_{Q_N}$ (MeV)}& \textbf{\boldmath$\sigma_L$ (MeV)}& \boldmath$r$\\[1pt]
			\toprule
		Skyrme~\cite{Dutra2012}		& 75.3	& 41.6	&178.5 & 27.2 	 & -0.843\\
		RMF~\cite{Dutra2014}		& -211.9	& 85.8			& 421.4 & 17.4 & -0.017\\
Combined~\cite{Dutra2012,Dutra2014}		& -53.4			& 61.2			& 341.2 & 32.1 & -0.498\\
Combined~\cite{Dutra2012,Dutra2014}	+ UGC/UGPC	& 7.86			& 58.2			& 297.1 & 21.6 & -0.378\\
		$\chi$EFT (SNM+PNM)		& -123.3			& 60.4			& 381.3 	& 8.1			& -0.686\\
$\chi$EFT (PNM)		& 112.8			& 51.9			& 90.7 	& 7.9			& 0.398\\

			\bottomrule
		\end{tabularx}
%	\end{adjustwidth}
%	\noindent{\footnotesize{\textsuperscript{1} This is a table footnote.}}
\end{table}

\begin{figure}[h]
%\hspace*{-1.8cm}\vspace*{-1.1cm}
\begin{adjustwidth}{-\extralength}{0cm}
\centering
\hspace*{-0.7cm}\includegraphics[width=10.5 cm,angle=180]{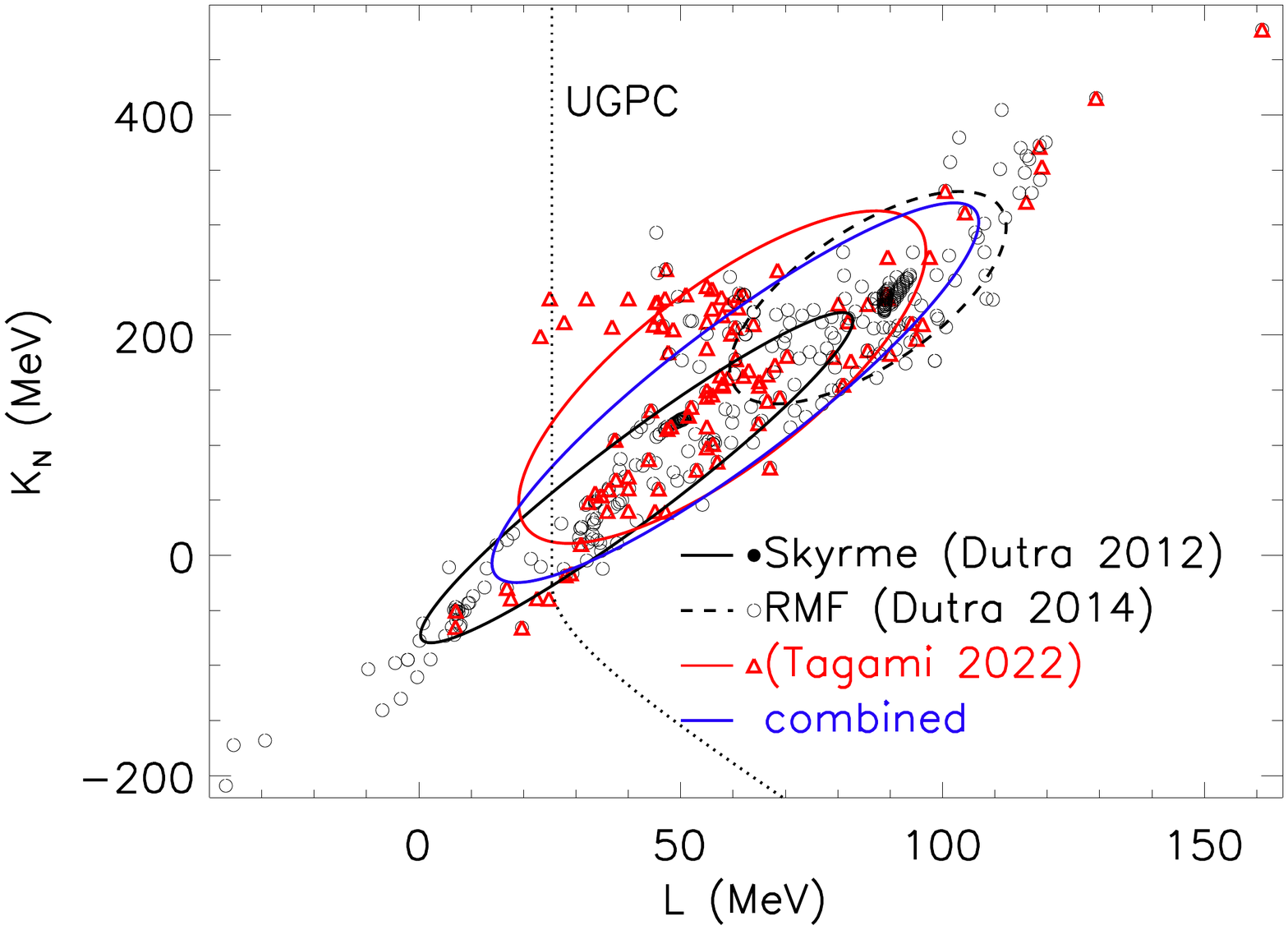}
\hspace*{-2cm}\includegraphics[width=10.55 cm,angle=180]{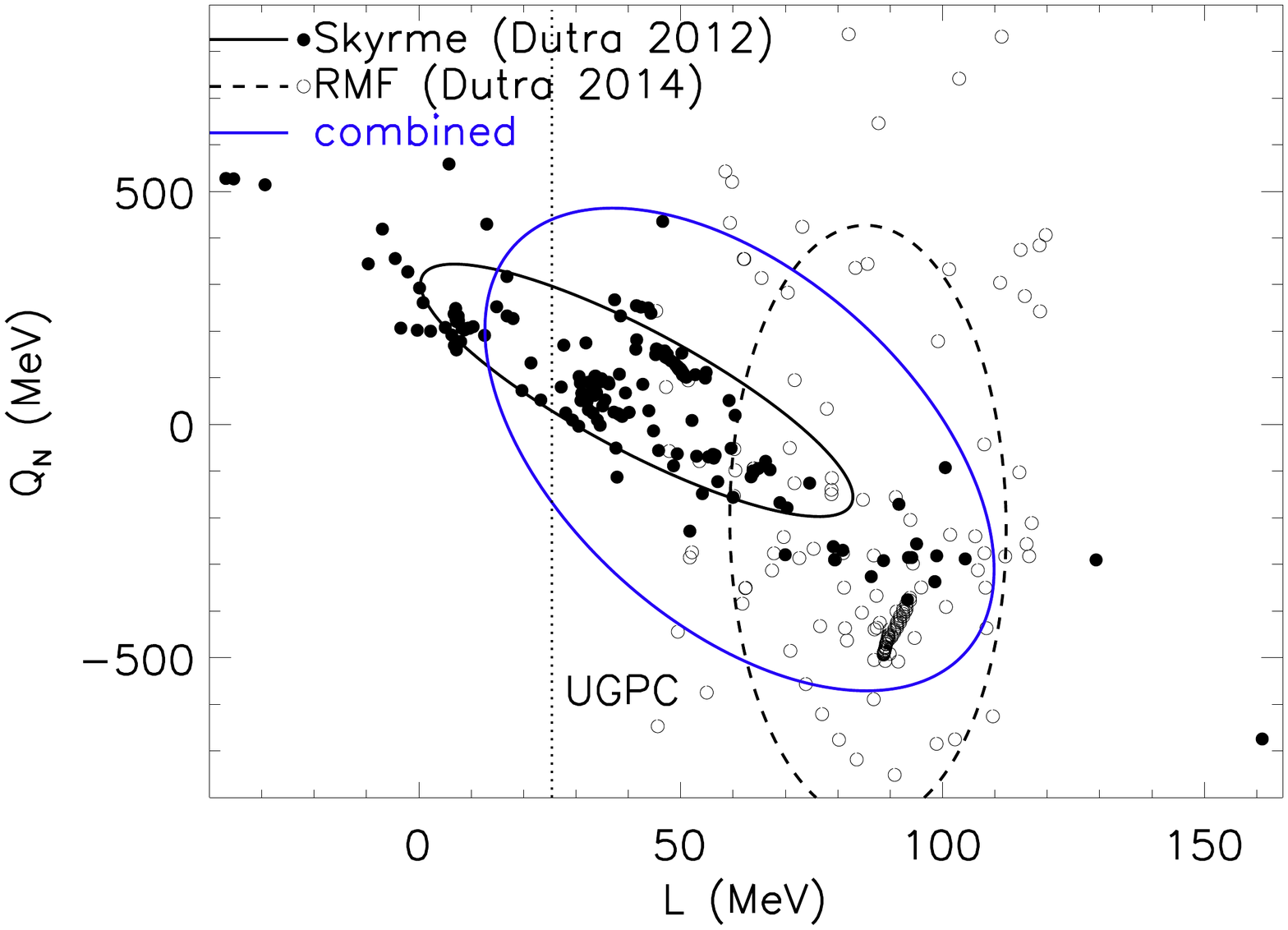}
\vspace*{-0.8cm}
\end{adjustwidth}
\caption{The same as Figure \ref{fig:correl0} but for correlations between $K_N$ and $L$ (\textbf{left panel}) and $Q_n$ and $L$ (\textbf{right panel}).  The Unitary Gas Pressure Conjecture restricts allowable regions to the right of the dotted lines labelled UGPC. \label{fig:correl1}}
\end{figure}

It is also possible to consider correlations involving incompressibilities and skewnesses, It will be seen to be more straightforward to consider $K_N=K_{1/2}+K_{sym}$ and $Q_N=Q_{1/2}+Q_{sym}$ rather than $K_{sym}$ and $Q_{sym}$.
Figure \ref{fig:correl1} displays these correlations for the interactions compiled by Refs.~\cite{Dutra2012,Dutra2014,Tagami2022}.  
Generally, it is seen that $K_N$ and $L$ are more highly correlated than $S_V$ and $L$ for model interactions, especially for Skyrme-like interactions.  Although $Q_N$ and $L$ are highly correlated for Skyrme-like interactions, they are much less correlated for RMF interactions.

It is clear there are systematic differences between the behaviors of Skyrme and RMF interactions.  In particular, Skyrme forces tend to display higher degrees of parameter correlations than do RMF forces.
In addition, mean values of the parameters $S_V, L$ and $K_N$ are larger, and $Q_{N}$ is smaller, for RMF compared to Skyrme forces. Importantly, these trends raise predicted values of neutron skin thicknesses of neutron-rich nuclei, and, further, raise values of $P_N$ for $n>n_s$ which increases estimated values of neutron star radii as shown in section 5.  For the compilations of Refs.~\cite{Dutra2012,Dutra2014}, there are 2 1/2 times as many surviving Skyrme  interactions as RMF interactions, but for the compilation of Ref.~\cite{Tagami2022}, the two types are more equally represented.  These relative populations are reflected in their combined correlation.

\begin{figure}[H]
\vspace*{-0.7cm}
\begin{adjustwidth}{-\extralength}{0cm}
\centering
\hspace*{-1.2cm}\includegraphics[width=10.5 cm,angle=180]{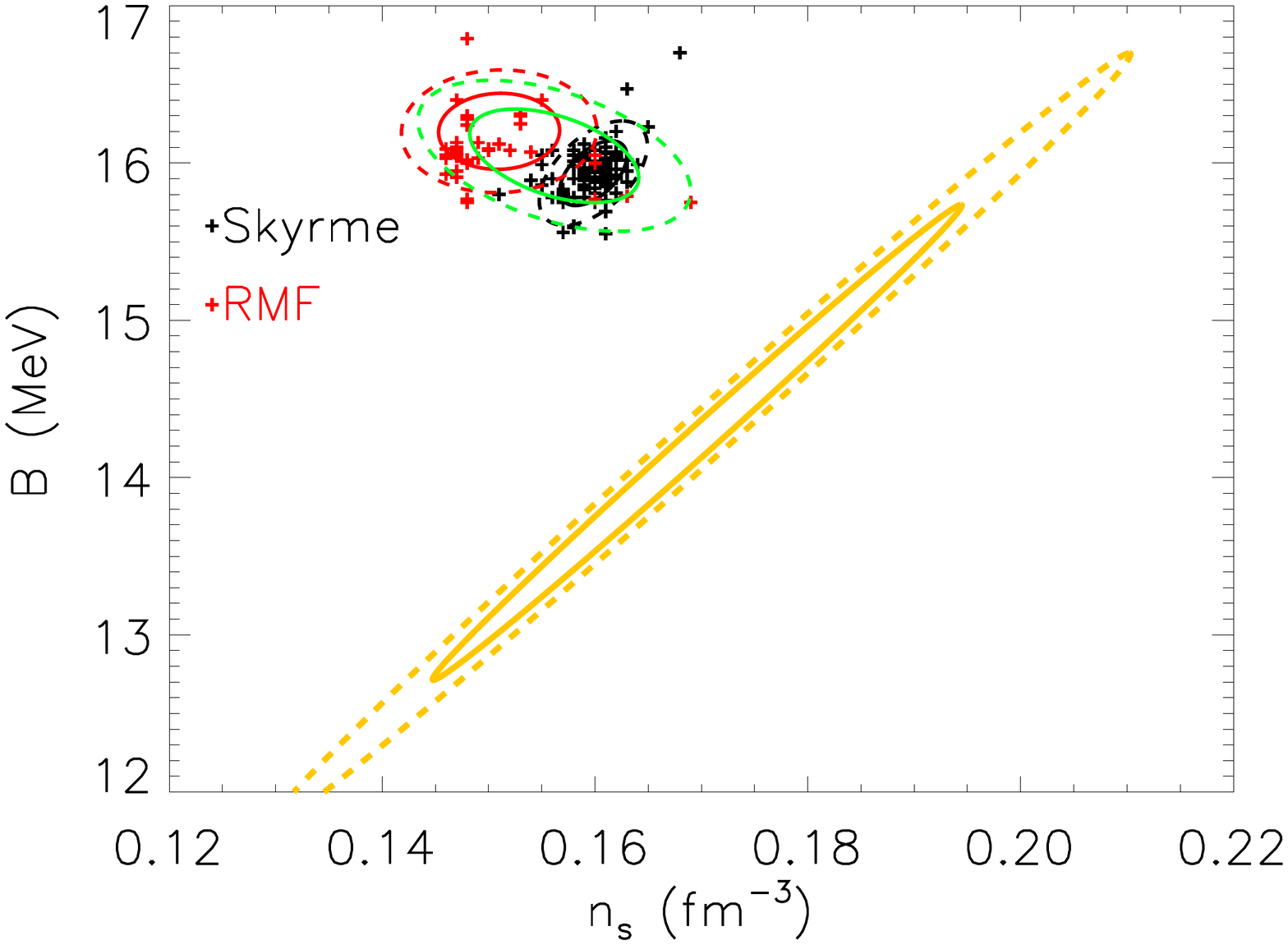}
\hspace*{-2cm}\includegraphics[width=10.5 cm,angle=180]{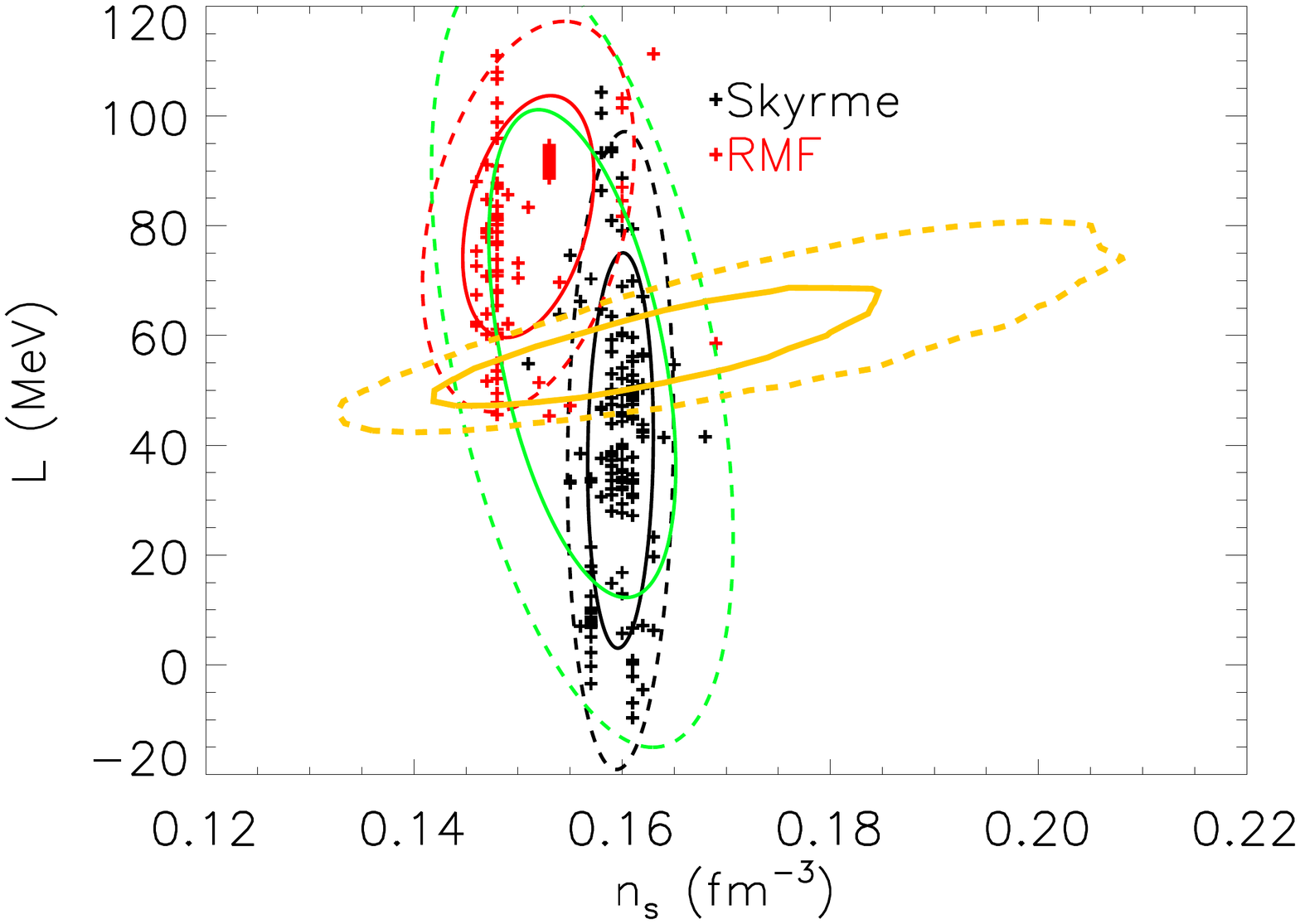}
\end{adjustwidth}
\vspace*{-0.5cm}
\caption{Correlations among symmetric energy parameters of forces in compilations of Refs.~\cite{Dutra2012,Dutra2014}.  The left and right panels show the $B-n_s$ and $L-n_s$ correlations, respectively.  Individual interactions are shown by filled circles.  68.3\% and 95\% confidence ellipses for Skyrme (RMF) interactions are shown by black (red) solid and dashed ellipses, respectively; green ellipses show the confidence ellipses for the combined force models.
68.3\% and 95.5\% confidence regions determined from $\chi$EFT calculations of SNM plus PNM (see Section~\ref{sec:NMT}) are shown by the orange solid and dotted curves, respectively.  
\label{fig:correl2}}
\end{figure}

\begin{figure}[H]
\begin{adjustwidth}{-\extralength}{0cm}
\centering
\hspace*{-0.7cm}\includegraphics[width=10.5 cm,angle=180]{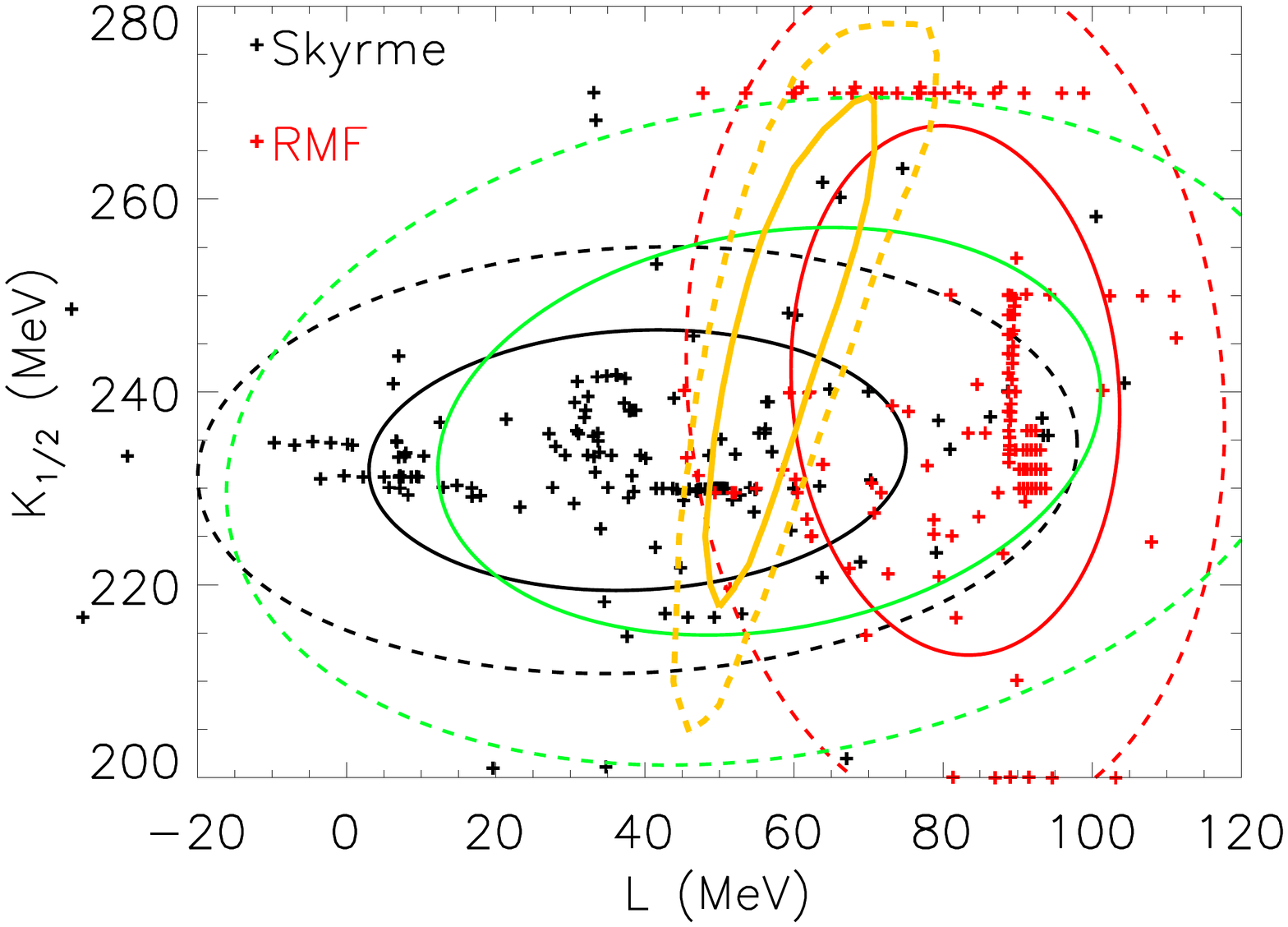}
\hspace*{-2cm}\includegraphics[width=10.5 cm,angle=180]{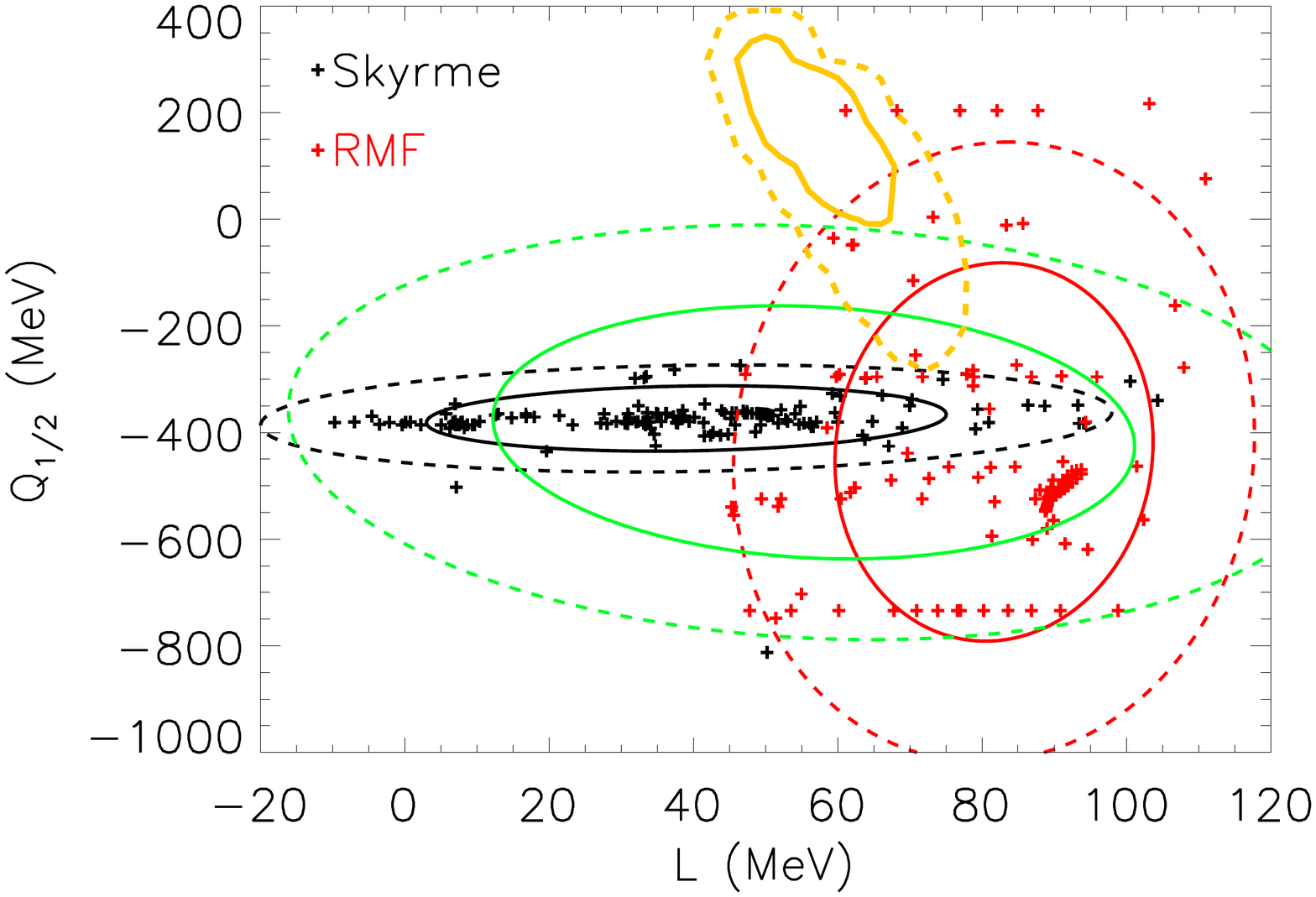}
\end{adjustwidth}
\vspace*{-0.5cm}
\caption{The same as Figure \ref{fig:correl2} except the left and right panels show the $L-K_{1/2}$ and $L-Q_{1/2}$ correlations, respectively.  \label{fig:correl3}}
\end{figure}   
Other systematic differences exist for symmetric matter parameters as well. For comparison, we display, using the databases of Refs.~\cite{Dutra2012,Dutra2014}, correlations among the symmetric energy parameters $n_s, B, K_{1/2}$, and $Q_{1/2}$, or between these symmetric energy parameters and $L$, in Figures~\ref{fig:correl2} and \ref{fig:correl3}.
%and their parameters in Table \ref{tab:symcor}.  
The parameters cluster in what defines the empirical saturation window, but display themselves display relatively weak correlations.  
   The UNEDF Collaboration~\cite{Kortelainen2010} also confirmed the lack of significant correlations among $B, n_s$ and $K_{1/2}$ and between those parameters and $L$ or $S_V$.

\subsection{Neutron Matter Theory\label{sec:NMT}}
A major recent advance in the understanding of nuclear matter has been made possible through the development of chiral effective field theory ($\chi$EFT)~\cite{Weinberg1967, Weinberg1968} which provides the only known framework allowing a systematic expansion of nuclear forces at low energies~\cite{Epelbaum2009,Machleidt2011,Hammer2020,Tews2020a} based on
the symmetries of quantum chromodynamics, the fundamental theory of the strong
interaction. In particular, $\chi$EFT allows one to derive systematic estimates of uncertainties
of thermodynamic quantities~\cite{Drischler2019,Leonhardt2020,Drischler2021a,Drischler2020b} for zero-temperature matter for densities up to $\sim2n_s$ with two- and three-nucleon interactions at the next-to-next-to-next-to-leading order (N3LO).
The energy and pressure of SNM are presented, for example by Ref.~\cite{Drischler2020}, as central values and their standard deviations as a function of density.   Corresponding values of the energy and pressure of PNM are also given.  

Ref.~\cite{Drischler2020} found that the energies and pressures, and the their uncertainties, for PNM and SNM are each significantly correlated, and also significantly correlated with each other.  From these calculations, Ref.~\cite{Drischler2020} thereby determined the correlation between $S_V$ and $L$ tabulated in Table~\ref{tab:svldata}. 
It is significantly flatter than from mass fitting, but has very consistent mean values.

\begin{figure}[h]
\vspace*{-.75cm}
\begin{adjustwidth}{-\extralength}{0cm}
\centering
\hspace*{-0.7cm}\includegraphics[width=10.5cm,angle=180]{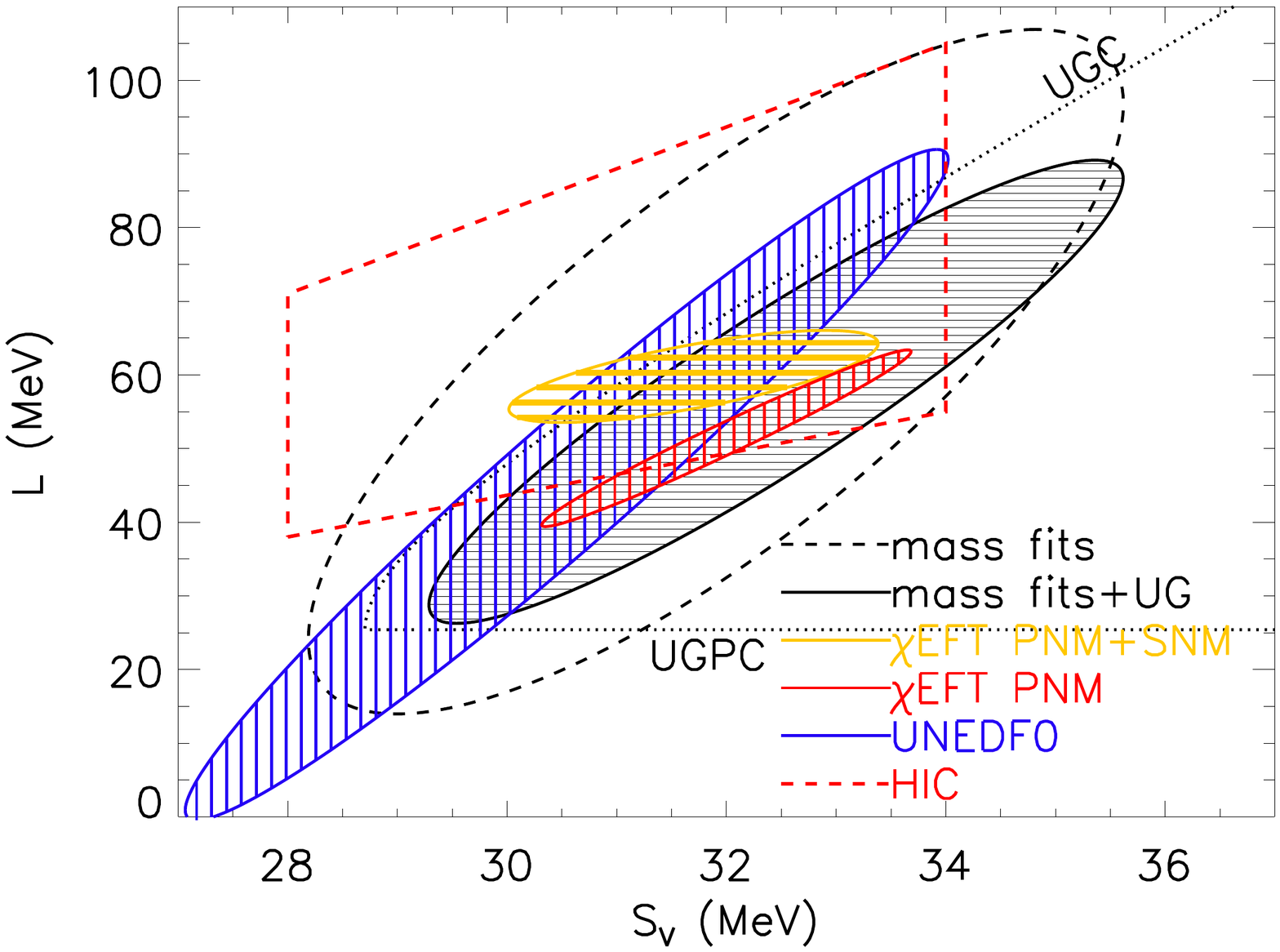}
\hspace*{-2cm}\includegraphics[width=10.5cm,angle=180]{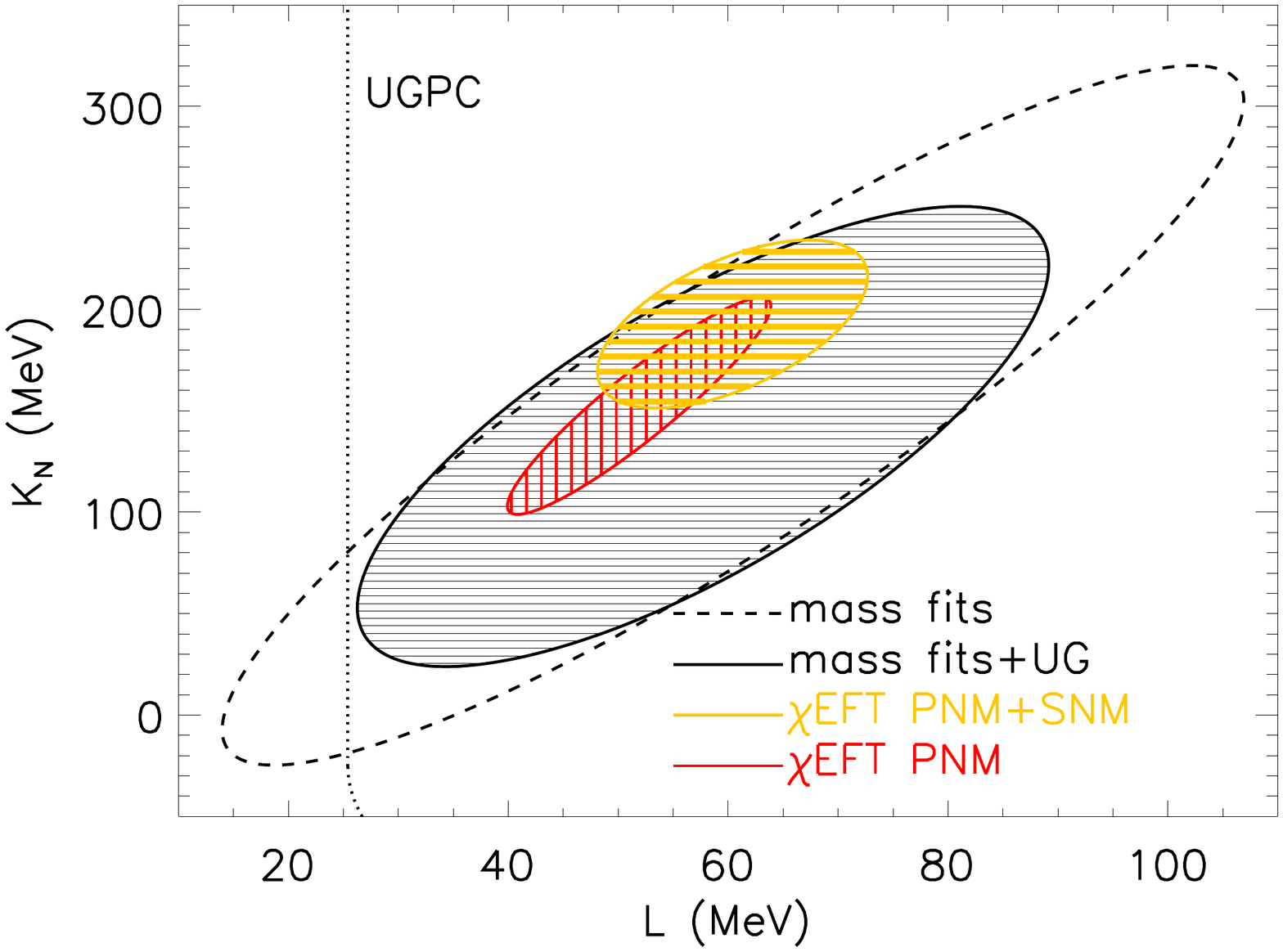}
\end{adjustwidth}
\vspace*{-1cm}
\caption{\label{fig:correl4}Black correlation ellipses for $S_V-L$ (\textbf{left panel}) and $K_N-L$ \textbf{(right panel)} use model interaction data~\cite{Dutra2012,Dutra2014,Tagami2022}, with (solid)  and without (dashed) application of Unitary Gas Constraints ~\cite{Tews2017} boundaries (dotted).  The blue confidence ellipse shows UNEDF~\cite{Kortelainen2010} results assuming $\sigma=1.2$ MeV.  The red (brown) confidence ellipses are from chiral EFT studies~\cite{Drischler2020} using PNM results with empirical saturation properties (combined PNM+SNM results).  The red-dashed quadrilateral are limits determined from elliptic flows in heavy-ion collisions~\cite{Tsang2009}.}
\end{figure}
These results can be generally reproduced directly using the SNM and PNM results and assuming a high degree of correlation between them.  The SNM calculations predict a distribution of saturation densities $n_s$ (defined by the relation $P_{1/2}(n_s)=0$, as well as distributions of binding energy ($B=-E_{1/2}(n_s)$), incompressibility ($K_{1/2}$) and skewness ($Q_{1/2}$) parameters.  These distributions are displayed in Figures~\ref{fig:correl2} and \ref{fig:correl3}.

Combining these results with PNM calculations and their distributions of neutron matter energy ($E_N(n_s)$) and pressures ($P_N(n_s)$) at the saturation density then yields distributions of $S_V$ and $L$ from $S_V=E_N(n_s)-B$ and $L=3P_N(n_s)/n_s$, shown in orange in Figure~\ref{fig:correl4} and tabulated in Table~\ref{tab:svldata}.  
This procedure gives a similar slope and mean parameter values as Ref.~\cite{Drischler2020}, but systematically larger standard deviations.  The differences may be due to our underestimate of correlations in and between SNM and PNM calculations.
The analysis can be extended to higher-order neutron matter parameters.  The case of $K_N$ is displayed in orange in Figure~\ref{fig:correl4}.

In contrast to mass fitting, the uncertainties of $B$ and $n_s$ from SNM $\chi$EFT calculations are extremely large and are also strongly correlated as seen in Figure \ref{fig:correl2}.  Most notably, the confidence ellipse does not pass near the empirical saturation window defined by Skyrme or RMF fits to nuclear properties.  Other symmetric energy parameters also have large uncertainties and show correlations (Figure~\ref{fig:correl3}).   Furthermore, the values of $Q_{1/2}$ are  inconsistent with those found from nuclear mass fits.

The failure of $\chi$EFT calculations of SNM to saturate inside the empirical saturation window, together with inconsistent values of $Q_{1/2}$, indicates that PNM calculations are much more reliable than SNM calculations at present.  This is not surprising, considering that the latter emerges from a delicate cancellation sensitive to the short- and intermediate-range three-body interactions at next-to-next-to-leading order, in contrast to PNM where these interactions are Pauli-blocked~\cite{HLPS10}.  

Therefore, we alternatively infer  symmetry energy parameters using only $\chi$EFT PNM results for the energy and pressure, including their standard deviations, but coupled with $n_s$ and $B$ values randomly chosen from within the empirical saturation window shown in Figure \ref{fig:correl2}.   This alternate $\chi$EFT $S_V-L$ correlation and those involving $K_N$ and $L$ are shown in Figure \ref{fig:correl4}.  Interestingly, the correlations so determined have confidence regions with central values consistent with those from mass fitting but with noticeably smaller uncertainties, greater degrees of significance, and also only slightly different slopes.
Furthermore, ranges of $S_V, L, K_N$ and $Q_N$ values are compatible with those observed from mass fits, neutron skin and dipole polarizability measurements, as well as astrophysical studies, as shown below. 

\subsection{The Unitary Gas Conjecture}
Ref.~\cite{Tews2017} proposed a constraint on the symmetry parameters arising from the conjecture that the energy of pure neutron matter was greater, at all densities, than that of a unitary gas.  A unitary gas is an idealized theoretical collection of fermions interacting only via pairwise $s$-wave interactions with an infinite scattering length and a vanishing effective range.  The average particle separation in such a gas is the only length scale, so the energy of the unitary gas, $E_{UG}$, is proportional to the Fermi energy,
\begin{equation}
E_{UG}={3\hbar^2k_F^2\over10m_N}\xi_0\simeq12.7\left({n\over n_s}\right)^{2/3}=E_{UG,0}u^{2/3},
\label{eq:ug}\end{equation}
where $k_F=(3\pi^2n_s)^{2/3}$ is the Fermi wave number at the saturation density, $m_N$ is the neutron mass, the Bertsch parameter, which is experimentally measured~\cite{Ku2012,Zurn2013} to be $\xi_0\simeq0.37$, and $E_{UG,0}\simeq12.6$ MeV.  In reality, pure neutron matter at low densities has finite scattering length and range, but both properties lead to larger energies than for a unitary gas.  In addition, three-body forces in neutron matter are known to be repulsive, further increasing its energy.  

The Unitary Gas Conjecture (UGC) states that $E_N\ge E_{UG}$ at all densities.  If it is minimally satisfied $E_N(u_t)=E_{UG}(u_t)$ at some arbitrary density $u_t$,  in order for it to remain satisfied at higher and lower densities requires~\cite{Tews2017}
\begin{equation}
\left({dE_N\over du}\right)_{u_t}=\left({dE_{UG}\over du}\right)_{u_t}.
\label{eq:UGC}\end{equation}
These conditions automatically impose constraints on the parameters $S_V$ and $L$ if the symmetry energy is expanded as in Equation~(\ref{eq:s}).  For example, it sets a minimum value $S_{V,min}=B+E_{UG,0}$ where $L=2E_{UG,0}$.  Further using the correlations shown in Figure~\ref{fig:correl1} to eliminate $K_{N}$ and $Q_{N}$, and using mean values for $\xi_0, n_s$ and $B$, the resulting constraint on $S_V$ and $L$ is displayed in Figure~\ref{fig:correl0} and in subsequent figures. This bound is relatively insensitive to assumed values of $K_N$ and $Q_N$~\cite{Tews2017}.  It is notable that the UGC is obeyed by nearly all $\chi$EFT results when $E_N$ and empirical values for $n_s$ and $B$ are used (PNM method), but only by about half of   $\chi$EFT results using both $E_N$ and $E_{1/2}$ (SNM+PNM method).  Similarly, most Skyrme interactions obey the UGC while most RMF forces do not. Even though the exact UGC boundary depends on uncertainties in $\xi_0, n_s, B$ and the $K_N-L$ and $Q_N-L$ correlations, it serves as a valuable consistency check.  It supports our previous argument that $\chi$EFT studies of symmetric matter are not presently accurate enough to provide significant constraints, and, further, that RMF forces do not seem to be as well-suited to fitting nuclear properties as are Skyrme forces.

Because the UGC establishes a lower limit to the energy of pure neutron matter, it effectively sets a lower limit to both the radius and tidal deformability of a neutron star as a function of its mass~\cite{Zhao2018}, being more restrictive than causality in this regard. 

Although the UGC cannot impose corresponding bounds in the $K_N-L$ or $Q_N-L$ planes, it is possible to propose a corollary Unitary Gas Pressure Conjecture (UGPC), which states that the neutron matter pressure is always greater than the unitary gas pressure for densities larger than $n_s$.  Comparisons shown by Ref.~\cite{Tews2017} show that this is the case for recent neutron matter calculations and true for PNM chiral EFT studies even considering 90\% lower confidence bounds.  For $u<1$, however, it is possible for the unitary gas pressure to be larger than the neutron matter pressure.  Using the 90\% upper confidence bound suggested by Figure \ref{fig:correl1} for $Q_N$ as a function of $L$, but ignoring the correlation between $K_N$ and $L$, the UGPC imposes the bound in the $K_N-L$ plane shown in Figure \ref{fig:correl1}.  All RMF forces obey the UGPC.  However, about 30\% of the formerly permitted Skyrme forces, those having  $L<2E_{UG,0}$ or $K_N<-2E_{UG,0}$, do not satisfy the UGPC.

One can tighten the constraints imposed from mass fitting by selecting only those forces whose values of $S_V$ and $L$ obey the UGC and the UGPC.  The parameters of the revised correlation ellipses for Skyrme and RMF forces are given  in Table \ref{tab:svldata} and also shown in Figure \ref{fig:correl4}.

%%%%%%%%%%%%%%%%%%%%%%%%%%%%%%%%%%%%%%%%%
\section{Neutron Skin Thickness Constraints}
It has long been known that the neutron skin thicknesses of neutron-rich nuclei, such as $^{48}$Ca and $^{208}$Pb, are highly dependent upon the symmetry parameter $L$ and more weakly dependent upon $S_V$.  For example, the liquid droplet model predicts that
the difference between the mean neutron and proton radii is~\cite{Myers1969}
\begin{equation}
t_{np}={2r_oI\over3}{S_S\over S_V}[{1+S_SA^{-1/3}/S_V}]^{-1}
\label{eq:tnp}\end{equation}
if Coulomb effects are ignored, where $r_o=(4\pi n_s/3)^{-1/3}$ and $I=(N-Z)/(N+Z)$.  The $S_S/S_V$ term indicates that the radius difference primarily depends on $L$, and therefore the symmetry energy slope $dS/dn$ and the neutron matter pressure at $n_s$.  However, the appearance of the last term
in Equation (\ref{eq:tnp}) implies that a stronger correlation of $t_{np}$ exists with the symmetry energy slope $dS/dn$ at a smaller density than $n_s$, namely about $2n_s/3$~\cite{Brown2000,Centelles2009} which can be viewed as a sort of average nuclear density.  This mimics the situation concerning nuclear binding energies.  Ref.~\cite{Brown2000} showed, in particular, that the neutron skin thickness of $^{208}$Pb, that is, the root-mean-square neutron-proton radius difference $r^{208}_{np}$, is linearly correlated with the neutron matter pressure (which is proportional to $n^2dS/dn$) most strongly at the density $n_1=0.10$ fm$^{-3}$:
\begin{equation}
r^{208}_{np}\simeq{\left(dS/dn\right)_{0.1}\over 882\pm32{\rm~MeV~fm}^{-2}}
\label{eq:rb}\end{equation}
We will define the symmetry energy slope as
\begin{equation}
\tilde L(n)\equiv3n{dS\over dn}
\label{eq:tildel}\end{equation}
so that $\tilde L(n_s)=L$.
Equation (\ref{eq:rb}) is then equivalent to $r_{np}\propto\tilde L_1\equiv\tilde L(n_1)$, an estimate later generalized by several authors as
\begin{equation}
r^{208}_{np}=\tilde a+\tilde b{\tilde L}_1,
\label{eq:TB}\end{equation}
with $\tilde a$ and $\tilde b$ coefficients given in Table \ref{tab:r-call}.  

\begin{table}[h]
\center\caption{Coefficients for the relation $r_{np}^{208}=\tilde a/{\rm fm}+\tilde b\tilde L_1/{\rm MeV}$ and inferred values of $\tilde L_1$ from the error-weighted mean experimental value $r_{np}^{208}=0.166\pm0.017$ fm. Also given are two estimated values of $\tilde L_1$ from neutron skin measurements of Sn isotopes.  $^\dagger$0.01 fm uncertainty introduced for consistency. \label{tab:r-call}}
\begin{tabularx}{\textwidth}{CCCCC}
\toprule
\textbf{Reference}&\boldmath$\tilde a$ &\boldmath$\tilde b$&\textbf{\boldmath$\tilde L_1$ (MeV)}\\
\midrule
\cite{Brown2000}&0.0& $0.00378\pm0.00014$&$43.9\pm4.8$ \\
\cite{Typel2001}& $0.00994\pm0.01000$ &0.0036&$43.4\pm5.5$\\
\cite{Xu2020}&$0.0101\pm0.01^\dagger$&0.00377&$41.4\pm5.2$\\
\cite{Steiner2005}&$0.0148\pm0.0100$&0.00414&3$6.5\pm4.8$\\
\cite{Reed2021}&$0.0590\pm0.0028$&0.00313&$34.2\pm10.5$\\

\cite{Zhang2013} &\multicolumn{2}{c}{Sn isotopes}&$42.9\pm4.1$\\ \cite{Xu2020} &\multicolumn{2}{c}{Sn isotopes}&$43.7\pm5.3$\\
\midrule
&\multicolumn{2}{c}{Error-weighted mean}&$41.7\pm2.0$\\
\bottomrule
\end{tabularx}
\end{table}\unskip

\begin{figure}[H]
\vspace*{-1.5cm}
\hspace{-1.8cm}\includegraphics[width=14 cm,angle=180]{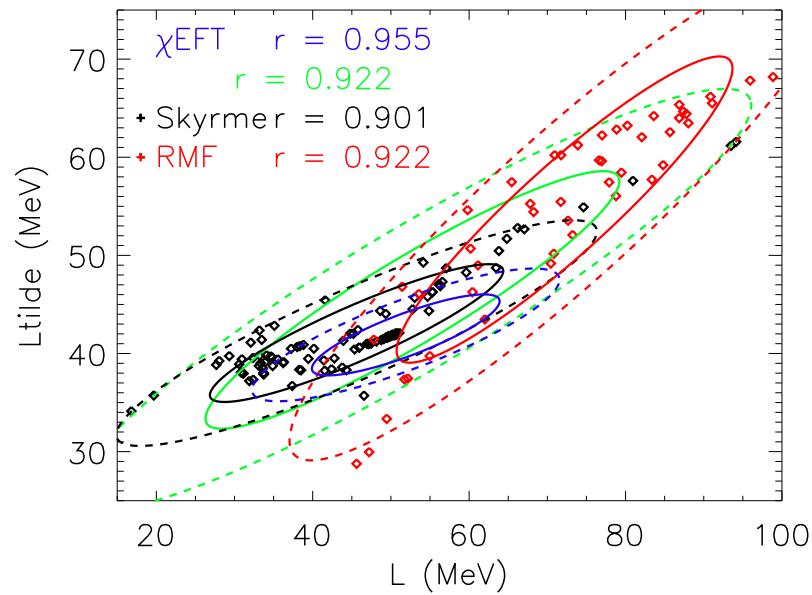}
\vspace*{-1.1cm}
\caption{Correlations between $\tilde L(u_1)$ (Ltilde) and $L$ for $\chi$EFT PNM (blue) and Skyrme (black) and RMF (red) interactions from Ref.~\cite{Dutra2012,Dutra2014}, respectively.  Green ellipses display the combined Skyrme and RMF correlations.  Correlation coefficients ($r$) are shown.    \label{fig:correl9}}
\end{figure}   

We note that $\tilde L$ evaluated at a subsaturation density can be related to other symmetry energy parameters through its Taylor expansion: 
\begin{equation}
\tilde L(u)=u\left[L+{K_{sym}\over3}(u-1)+{Q_{sym}\over18}(u-1)^2+\cdots\right].
\label{eq:tildel1}
\end{equation}
Given the strong correlations between $K_{sym}$ and $L$, and moderate correlations between $Q_{sym}$ and $L$, displayed for the Skyrme~\cite{Dutra2012} and RMF~\cite{Dutra2014} forces (both shown in Figure~\ref{fig:correl3}), as well as for $\chi$EFT for PNM using $B$ and $n_s$ from the empirical saturation window, it is not surprising that a strong correlation exists also between $\tilde L(u_1)$ and $L$ (Figure~\ref{fig:correl9}).    When the $K_{sym}-Q_{sym}-L$ correlations are combined in Equation ~(\ref{eq:tildel1}), they give a $1\sigma$ confidence ellipse centered at $L=50.09$ MeV and $\tilde L=41.86$ MeV, with standard deviations $\sigma_L=6.81$ MeV and $\sigma_{\tilde L}=2.55$ MeV and correlation 0.843.  The result $\tilde L=41.7\pm1.9$ MeV from the average Pb and Sn neutron skin thicknesses shown in Table~\ref{tab:r-call}) has nearly the same value and uncertainty as those from mass fitting and neutron matter theory shown in Figure~\ref{fig:correl9}.   Their mean value of about 40 MeV is supported by the results of \mbox{Refs. ~\cite{Xu2020,Zhang2013}} who argued that a linear $r_{np}-\tilde L$ correlation also exists for neutron-rich Sn isotopes and derived similar values of $\tilde L$ near the density $2n_s/3$ from experimental data (Table ~\ref{tab:r-call}).  Collectively, these Pb and Sn studies yield an average value $\tilde L_1\simeq42$ MeV. The implied value $L=49.9\pm5.0$ MeV is therefore also in remarkable agreement.

\begin{figure}[H]
\vspace*{-1.3cm}
\hspace{-.5cm}
    \includegraphics[width=13cm,angle=180]{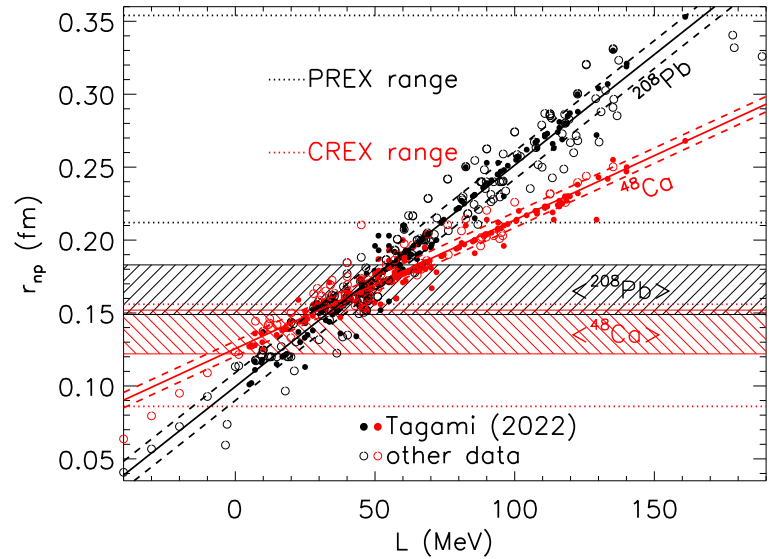}
  \vspace*{-1.1cm}  \caption{Neutron skin thicknesses of $^{48}$Ca (red) and $^{208}$Pb (black) from interactions compiled in Ref.~\cite{Tagami2022} (filled circles) and Refs.~\cite{Brown2017,Adhikari2022,Horowitz2014,Piekarewicz2012} (open circles).  Means (1 standard deviations) of linear correlations are shown as solid (dashed) lines.  The horizontal shaded bands indicate the 1 standard deviation ranges of the averaged experimental results.  The dotted black (red) lines indicate the 1 standard deviation range of $r_{np}^{208}$ ($r_{np}^{48}$) from PREX I+II~\cite{Adhikari2021} (CREX~\cite{Adhikari2022}).
    \label{fig:skin}}
\end{figure}

The pseudo-linear correlation between $\tilde L$ and $L$ implies good linear correlations should exist between $r_{np}$ with $L$ for both Pb and Sn nuclei.  Theoretical modeling, both from mean-field analyses and the dispersive optical model, supports an extension to $^{48}$Ca.  Linear relations are indeed validated by examining recent compilations~\cite{Tagami2022,Adhikari2022,Brown2017,Reed2021,Furnstahl2002,Reinhard2021} of neutron skin thicknesses for $^{208}$Pb and $^{48}$Ca using a multitude of both non-relativistic and relativistic interactions. The compilation from Ref.~\cite{Tagami2022} is especially notable in containing results from 206 Skyrme-like and RMF forces, and the other compilations contribute more than 200 additional values.  The model estimates from these compilations are displayed in Figure~\ref{fig:skin} and slopes, intercepts and standard deviations of the linear fits from each reference, as well as their overall means, are provided in Table \ref{tab:skinfit}.  
   The mean values of the skin thicknesses from all models shown are nearly the same, $r_{np}^{208}=0.19\pm0.05$ fm and $r_{np}^{48}=0.18\pm0.03$ fm, but the two correlations have different slopes.  Both are within $\pm1\sigma$ of the respective mean experimental measurements, see Tables \ref{tab:skin} and \ref{tab:skinCa}, indicating an overall consistency exists between theory and experiment even though most of the models displayed were not explicitly fit to neutron skin values but rather to binding energy data. In other words, there is no reason to expect that either conventional interactions or modeling lead to large systematic uncertainties with respect to calculations of neutron skin thicknesses.

\begin{table}[H]
\center\caption{Slopes, intercepts, and standard deviations of linear fits $r_{np}/{\rm fm}=a\pm\Delta a+bL/{\rm MeV}$.
\label{tab:skinfit}}
\begin{tabularx}{\textwidth}{CCC}
\toprule
\textbf{Reference}&\boldmath$a$&\boldmath$b$\\
\midrule
\multicolumn{3}{c}{$^{208}$Pb}\\
\midrule
\cite{Tagami2022}&$0.0963\pm0.0041$&0.001566\\
\cite{Adhikari2022}&$0.1028\pm0.0115$&0.001617\\
\cite{Brown2017}&$0.0964\pm0.0039$&0.001563\\
\cite{Reed2021}&$0.0865\pm0.0124$&0.001837\\
\cite{Furnstahl2002}&$0.0967\pm0.001447$&0.00145\\
\cite{Reinhard2021}&$0.0986\pm0.0137$&0.001537\\
\midrule
Mean&$0.0996\pm0.0096$&0.001518\\
\midrule
\multicolumn{3}{c}{$^{48}$Ca}\\
\midrule
\cite{Tagami2022}&$0.1250\pm0.0028$&0.000873\\
\cite{Adhikari2022}&$0.1261\pm0.0056$&0.000990\\
\cite{Brown2017}&$0.1290\pm0.0037$&0.000791\\
\midrule
Mean&$0.1255\pm0.0052$&0.000882\\
\bottomrule
\end{tabularx}
\end{table}\unskip

Nevertheless, it is worth mentioning that two recent more sophisticated theoretical predictions give divergent views.  Relatively small values of neutron skins are implied by coupled cluster ab-initio calculations which predict $r_{np}^{48}=0.135\pm0.015$ fm~\cite{Hagen2016} and  $r_{np}^{208}=0.17\pm0.03$ fm~\cite{Hu2021}, which are close to the respective experimental means (Table~\ref{tab:skin}).  On the other hand, the nonlocal dispersive optical model predicts that finite-size effects play an important role in enhancing the neutron skin, giving $r_{np}^{208}=0.25\pm0.05$ fm~\cite{Atkinson2020} and $r_{np}^{48}=0.249\pm0.023$ fm~\cite{Mahzoon2017}, both of which are considerably larger than the respective experimental means. 

The strong linear correlations existing between model calculations of the neutron skins of these two nuclei with $L$ obviously implies a strong linear correlation exists between the skin thicknesses themselves.   Neutron skin data from Ref.~\cite{Tagami2022} alone follows the linear relation
    \begin{equation}
r_{np}^{48}=( 0.0716\pm0.0006){\rm~fm}+0.5554r_{np}^{208}   \label{eq:capbt}
\end{equation}
with a very small uncertainty.  Including  179 skin calculations using additional Skyrme and RMF forces by Refs.~\cite{Piekarewicz2012, Horowitz2014,Brown2017,Adhikari2022}, the mean linear relation becomes
\begin{equation}
    r_{np}^{48}=(0.0730\pm0.0048){\rm~fm}+0.56157r_{np}^{208},
    \label{eq:capb}
    \end{equation}
which is virtually identical except for having a larger standard deviation reflecting a greater 
variation in underlying forces (and possibly less strict constraints regarding fitting nuclear
masses and charge radii). The $\pm1\sigma$ confidence bounds on the overall linear correlation are shown as straight dashed lines in Figure 9. We note that these references 
contain a number of the same interactions. These duplicate calculations show variations 
of as much as 0.01 fm, which should be included as a systematic modeling uncertainty. 
However, this uncertainty is small enough that it does not affect the results significantly. We also note that Ref. \cite{Reinhard2013} has argued that an accurate determination of
$r_{np}^{208}$ is insufficient to constrain $r_{np}^{48}$ because of the significant difference in
the surface-to-volume ratio of these two nuclei, a conclusion, however, 
not supported by our results.

\subsection{Neutron Skin Measurements and Correlations}
\begin{figure}[H]
\vspace*{-1.35cm}\includegraphics[width=11.5cm,angle=180]{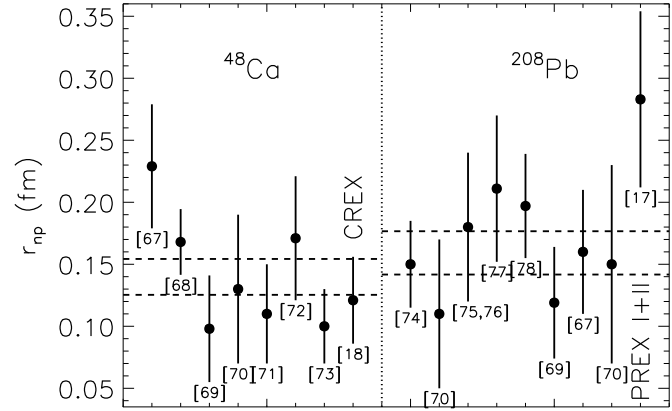}% Here is how to import EPS art
\vspace*{-1.25cm}
\caption{Neutron skin measurements~\cite{Ray1979,Zenhiro2018,Clark2003,Friedman2012,Gibbs1992,Gils1984,Shlomo1979,Adhikari2021,Tarbert2014,Klos2007,Brown2007,Zenhiro2010,Starodubsky1994,Adhikari2022} with 68\% confidence intervals and citations.  Horizontal dashed lines denote $\pm1$ standard deviations from the weighted means of experiments other than CREX or PREX I+II.\label{fig:measskin}}
\end{figure}

The nuclides $^{48}$Ca and $^{208}$Pb are especially important because they are the only stable neutron-rich, closed-shell, nuclei.  Measurements of their neutron skin thicknesses are summarized in Tables \ref{tab:skin} and \ref{tab:skinCa}, as well as Figure \ref{fig:measskin}.  The error-weighted mean of all tabulated experimental measurements is $r_{np}^{208}=0.166\pm0.017$ fm,  which is consistent with the average theoretical estimate \mbox{$r_{np}^{208}=0.170\pm0.008$ fm}.  The mean of historical measurements not including PREX is $r_{np}^{208}=0.159\pm0.017$ fm.  For $^{48}$Ca, the mean of all measurements is $r_{np}^{48}=0.137\pm0.015$ fm, about $2\sigma$ smaller than the average  theoretical estimate \mbox{$r_{np}^{48}=0.17\pm0.03$ fm}.  The average of historical measurements not including CREX is $r_{np}^{48}=0.140\pm0.017$ fm. 

\begin{table}[H]
\caption{$^{208}$Pb  neutron skin measurements and theoretical predictions with $1\sigma$ uncertainties
\label{tab:skin}}
%\vspace*{-.5cm}
%\center
\begin{tabularx}{\textwidth}{cCC}
\toprule
\textbf{\boldmath$^{208}$Pb Experiment}&\textbf{Reference}&$r_{np}^{208}$ (fm)\\
\midrule
&&\\[-8pt]
Coherent $\pi^0\gamma$ production&\cite{Tarbert2014}&$0.15^{+0.03}_{-0.04}$ \\
Pionic atoms&\cite{Friedman2012}&$0.15\pm0.08$\\
Pion scattering&\cite{Friedman2012}& $0.11\pm0.06$\\
$\bar p$ annihilation&\cite{Klos2007,Brown2007}&$0.18\pm0.06$\\
Elastic polarized $p$ scattering&\cite{Ray1979}&$0.16\pm0.05$\\
Elastic polarized $p$ scattering&\cite{Zenhiro2010}&$0.211^{+0.054}_{-0.063}$\\
Elastic $p$ scattering&\cite{Starodubsky1994}&$0.197\pm0.042$\\
Elastic $p$ scattering&\cite{Clark2003}&$0.119\pm0.045$\\
Parity-violating $e^-$ scattering (PREX I+II)&\cite{Adhikari2021}&$0.283\pm0.071$\\
\midrule
$^{208}$Pb experimental weighted mean&&$0.166\pm0.017$ \\
\midrule
Pygmy dipole resonances&\cite{Klimkiewicz2007}&$0.180\pm0.035$\\
$r_{np}^{\rm Sn}$ &\cite{Chen2010}&$0.175\pm0.020$\\
Anti-analog giant dipole resonance&\cite{Yasuda2014}&$0.216\pm0.048$\\
Symmetry energy $^{208}$Pb&\cite{Dong2015}&$0.158\pm0.014$\\
Dispersive optical model&\cite{Pruitt2020}&$0.18^{+0.25}_{-0.12}$\\
Dispersive optical model&\cite{Atkinson2020}&$0.25\pm0.05$\\
Coupled cluster expansion&\cite{Hu2021}&$0.17\pm0.03$\\
$r_{np}^{48}$ &\cite{Horowitz2014,Piekarewicz2012}, this paper&$0.128\pm0.040$\\
$\alpha_D^{208}$ &\cite{Reinhard2021}, this paper&$0.154\pm0.019$\\
$\alpha_D^{208}$ &\cite{Piekarewicz2012,Reed2021}, this paper&$0.188\pm0.017$\\
\midrule
$^{208}$Pb theoretical weighted mean&&$0.170\pm0.008$\\
\bottomrule
\end{tabularx}
\end{table}\unskip

\begin{table}[H]
\caption{$^{48}$Ca  neutron skin measurements and theoretical predictions with $1\sigma$ uncertainties. $^*$Uncertainty scaled upwards as per Ref.~\cite{Zyla2020}.
\label{tab:skinCa}}
\begin{tabularx}{\textwidth}{cCC}
\toprule
\textbf{$\boldmath^{48}$Ca Experiment}&\textbf{Reference}&\textbf{\boldmath$r_{np}^{48}$ (fm)}\\
\midrule
Elastic polarized $p$ scattering&\cite{Ray1979}&$0.229\pm0.050$\\
Elastic $p$ scattering&\cite{Shlomo1979}&$0.10\pm0.03$\\
Elastic $p$ scattering&\cite{Clark2003}&$0.098\pm0.043$\\
Elastic $p$ scattering&\cite{Zenhiro2018}&$0.168^{+0.025}_{-0.028}$\\
Pionic atoms&\cite{Friedman2012}&$0.13\pm0.06$\\
Pion scattering &\cite{Gibbs1992}&$0.11\pm0.04$\\
$\alpha$ scattering &\cite{Gils1984}&$0.171\pm0.050$\\
Parity-violating $e^-$ scattering (CREX)&\cite{Adhikari2022}&$0.121\pm0.035$\\
\midrule
$^{48}$Ca experimental weighted mean&&$0.137\pm0.015^*$\\
\midrule
Coupled-cluster expansion&\cite{Hagen2016}&$0.135\pm0.015$\\
Dispersive optical model&\cite{Mahzoon2017}&$0.249\pm0.023$\\
$r_{np}^{208}$&\cite{Horowitz2014}, this paper&$0.173\pm0.018$\\
\midrule
$^{48}$Ca theoretical weighted mean&&$0.17\pm0.03^*$\\
\bottomrule
\end{tabularx}
\end{table}\unskip

\subsection{Parity-Violating Electron Scattering Measurements}
A lot of attention has been paid to the recent PREX~\cite{Adhikari2021} and CREX~\cite{Adhikari2022}  measurements of the neutron skins of $^{208}$Pb and $^{48}$Ca, respectively, using parity-violating electron scattering, which is claimed to have less modeling systematic uncertainty than other experimental methods~\cite{Thiel2019}. Interestingly, the PREX measurement is more than 1 standard deviation higher than the mean value of previous $^{208}$Pb experiments, while the CREX measurement is smaller than the mean of previous $^{48}$ Ca experiments, but by less than 1 standard deviation.  Fitting the neutron skin thickness from parity-violating scattering of either nuclide alone would give vastly different values for $L$, about 110 MeV for Pb~\cite{Reed2021} and 0 MeV for Ca, as can be seen by reference to Figure~\ref{fig:skin}. Even using the mean values of all the experimental results would produce disparate values of $L$, about 40 MeV and 10 MeV, respectively, although they would lie within a standard deviation of each other.  It is important to note that the weighted mean experimental value of $r_{np}^{208}$  decreases  by only about 0.007 fm and that of $r_{np}^{48}$ increases by only about 0.003 fm when the PREX and CREX results are excluded.  Without  compelling reasons to favor measurements of either nuclide, our approach is to instead attempt to simultaneously satisfy experimental information for both nuclides.  

We follow two strategies for satisfaction of joint Ca-Pb measurements.  First, one could take the approach that the PREX and CREX experiments  qualitatively have fewer systematic uncertainties than other approaches, and only use those measurements.  Alternatively, an agnostic approach would be to instead consider the weighted means of all measurements.
 
\begin{figure}[H]
\vspace*{-1.3cm}
\hspace*{-.5cm}
\includegraphics[width=13cm,angle=180]{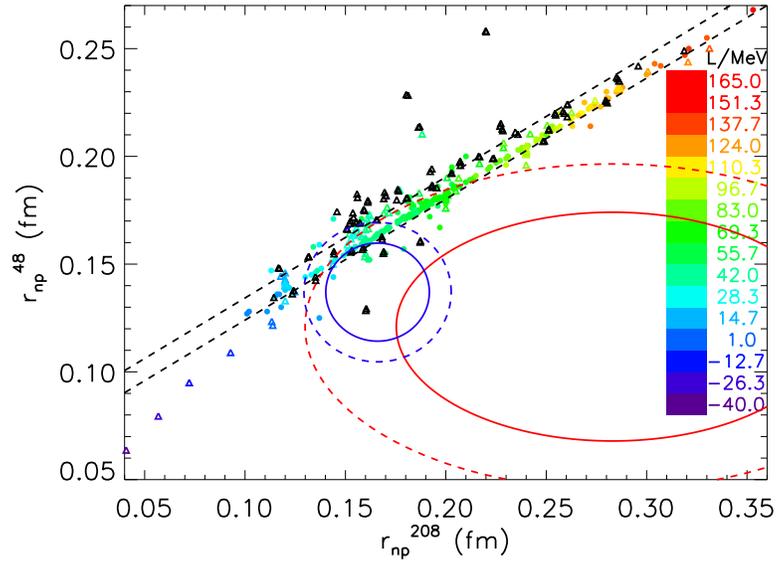}
\vspace*{-1cm}  \caption{Neutron skin thicknesses of $^{48}$Ca and $^{208}$Pb from interactions compiled in Ref.~\cite{Tagami2022} (filled circles) and Refs.~\cite{Brown2017, Adhikari2022,Horowitz2014,Piekarewicz2012} (triangles).  Colors indicate $L$ values where known; black triangles indicate points where $L$ values are unspecified.  Standard deviations of a linear correlation Equation~(\ref{eq:capb}) are shown as dashed lines.  The red (blue) confidence ellipses are from PREX I+II~\cite{Adhikari2021} and CREX~\cite{Adhikari2022} (mean of all experiments); solid (dashed) ellipses are 68\% (90\%) confidence. 
    \label{fig:capb}}
\end{figure}

As can be seen in Figure \ref{fig:capb}, the PREX I+II value for $r_{np}^{208}$ is too large and the CREX value for $r_{np}^{48}$ is too small to permit any of the reference interactions from the compilation of Refs.~\cite{Tagami2022,Adhikari2022,Brown2017,Horowitz2014,Piekarewicz2012} from satisfying both of them to within 68\% confidence.  The situation is different when considering the mean experimental results for $^{208}$Pb and $^{48}$Ca, with 4\% of the reference interactions simultaneously satisfying them to 68\% confidence. 

  A much larger number of interactions satisfy skin thickness measurements for both nuclei when considering 90\% confidence regions. About 40\% of the interactions from Ref.~\cite{Tagami2022} can simultaneously satisfy the UGC, UGPC, PREX I+II and CREX results, and these have 0 MeV $<L<72$ MeV.  
  Similarly, about 24\% of these interactions simultaneously lie within the 90\%  confidence region of the averages of all experiments, and also satisfy the UGC and UGPC, and these have 0 MeV $<L<58$ MeV.

\begin{figure}[H]
\vspace*{-1cm}
\begin{adjustwidth}{-\extralength}{0cm}
\centering
\hspace*{-0.7cm}\includegraphics[width=10.5cm,angle=180]{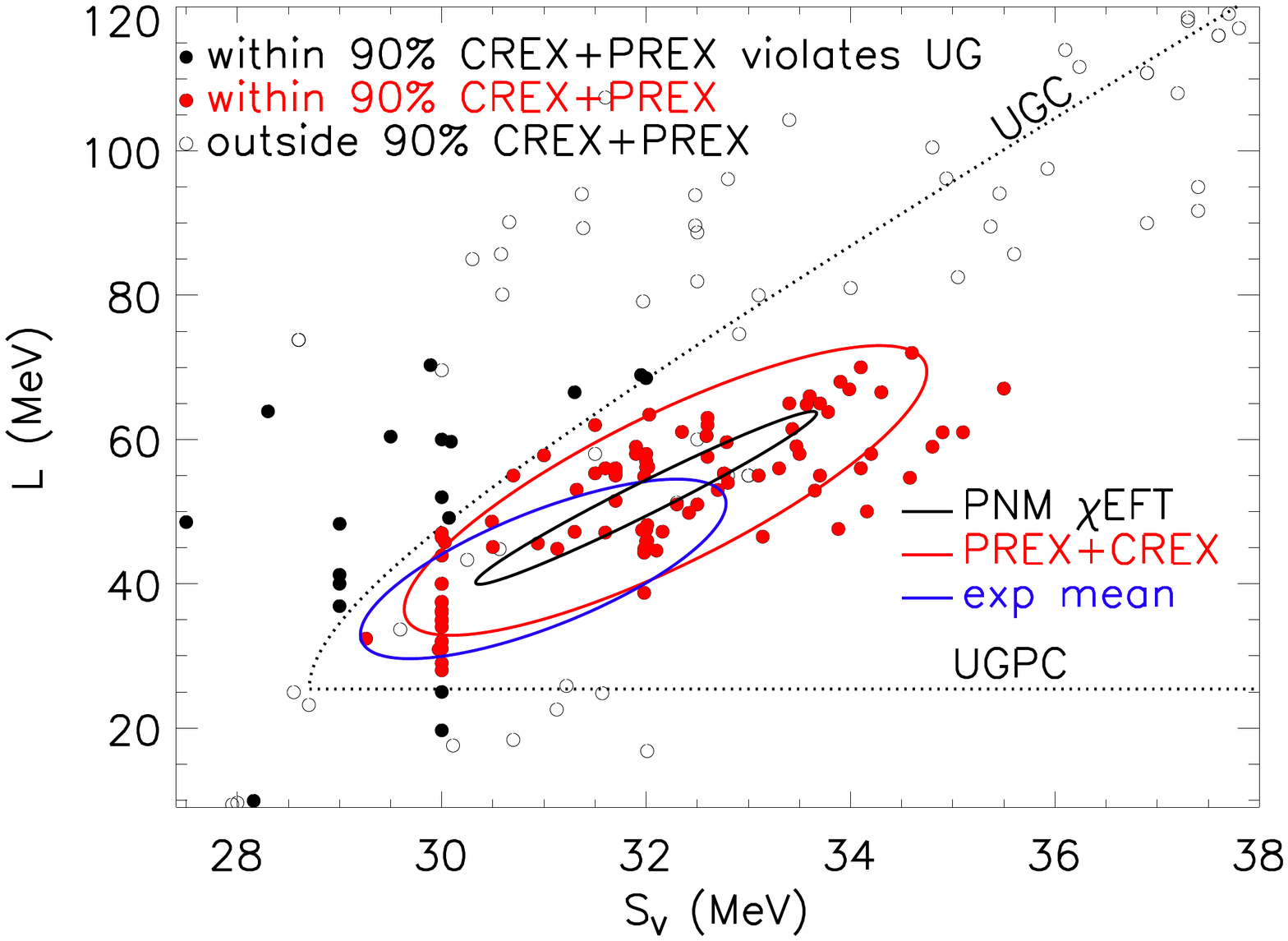}
\hspace*{-2cm}\includegraphics[width=10.5cm,angle=180]{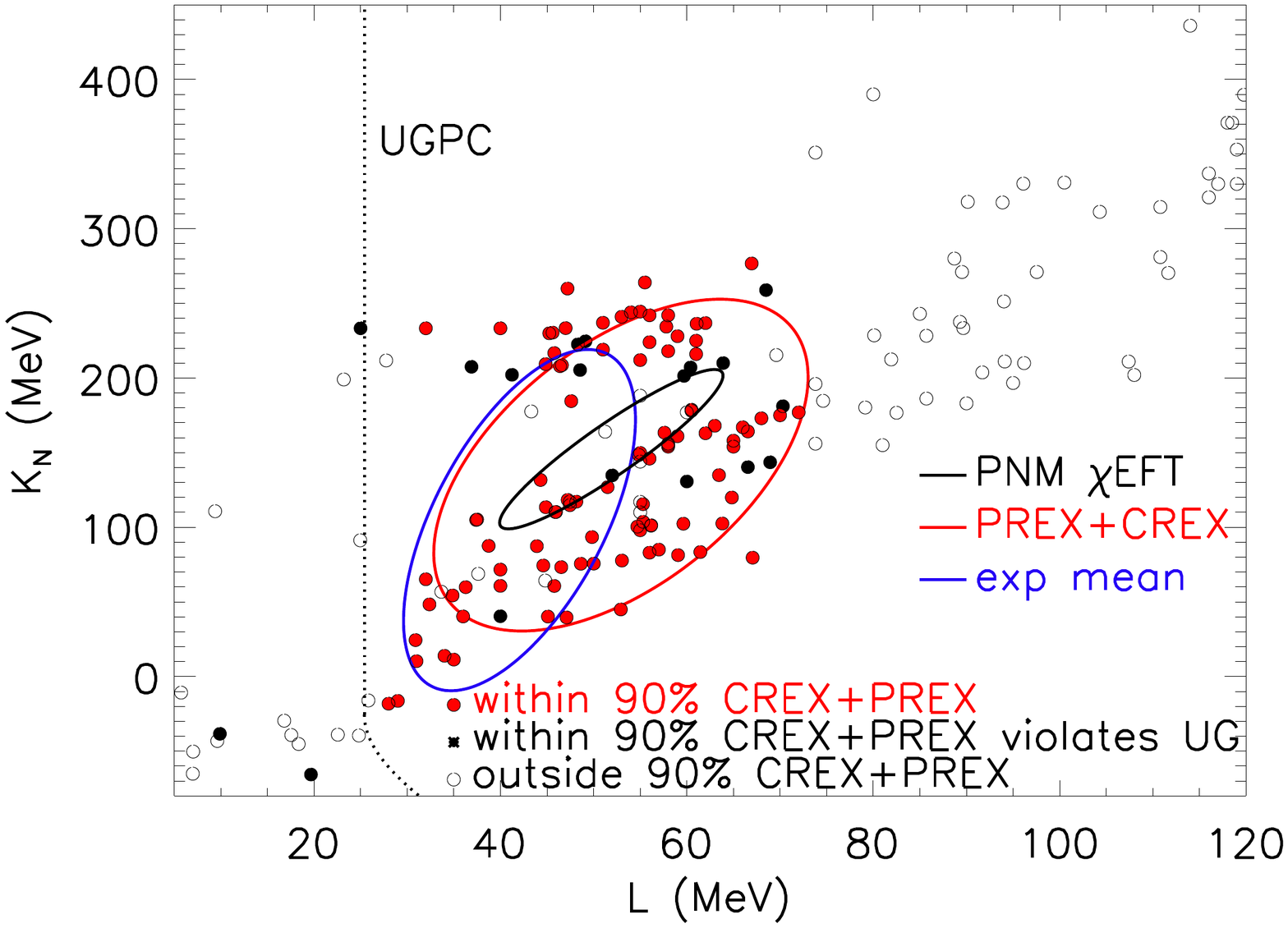}% Here is how to import EPS art
\end{adjustwidth}
\vspace*{-1cm}
\caption{Symmetry parameters $S_V-L$ (\textbf{left panel}) and $K_N-L$ (\textbf{right panel}) jointly satisfying parity-violating experiments to within (exceeding) 90\% confidence are shown as red and black filled (black open) circles; filled black circles violate Unitary Gas constraints (dotted boundary).  Red (blue) confidence ellipses are for models satisfying Unitary Gas constraints weighted by their two-dimensional Gaussian probability defined by the parity-violating (red) and average (blue) experimental $r_{np}^{48}$ and $r_{np}^{208}$ measurements and uncertainties.  The black confidence ellipse shows PNM $\chi$EFT results.}
\label{fig:capbsvl}\end{figure}

The associated permitted region in $S_V-L$ space can be ascertained by weighting those interactions~\cite{Tagami2022} satisfying both unitary gas constraints, and which also have tabulated $S_V, L, r_{np}^{48}$ and $r_{np}^{208}$ values, 
by their probabilities given by a
two-dimensional Gaussian defined by the skin measurements and their uncertainties.  Results are shown in Figure~\ref{fig:capbsvl} and tabulated in Table~\ref{tab:svldata} for both approaches.  Interestingly, using CREX+PREX measurements to define the $r_{np}^{48}-r_{np}^{208}$  probabilities gives a confidence ellipse in substantial agreement with the PNM $\chi$EFT result.  Using the mean of other skin measurements gives somewhat smaller mean values of $S_V$ and $L$. 

It is important to note that this internal consistency among neutron skin measurements, mass fitting and neutron matter theory, using either approach, is not particularly sensitive to whether relativistic or non-relativistic interactions are considered, suggesting it is relatively free of associated systematic uncertainties.  Partly, this is due to the moderate values of $L$ that are favored, eliminating most RMF interactions in compilations.

We compare to 
 Ref.~\cite{Danielewicz2014}, who combined data from isobaric analog states and the mean of experimental Pb neutron skin measurements (taken to be to $r_{np}^{208}=0.159\pm0.041$ fm) to infer $L\simeq50\pm12$ to 68\% confidence. Also, Ref.~\cite{Furnstahl2002} used experimental values for Pb to inferr $19\simle L/{\rm MeV}\simle77$.  However, neither of these sources considered the neutron skin of $^{48}$Ca.

We note an inference of $S_V=30.2^{+4.1}_{-3.0}$ MeV and $L=15.3^{46.8}_{-41.5}$ MeV using a combined analysis of PREX and CREX measurements was recently obtained by Ref.~\cite{Zhang2022}.  Their method differs from the present analysis in that they optimized parameters for a Skyrme-like interaction using weak and charge form factors from PREX and CREX rather than inferred skin thicknesses, which are subject to additional theoretical uncertainty, as well as binding, breathing mode and neutron-proton Fermi energies, and charge and diffraction radii.  Their estimates are compatible with the present results but have larger uncertainties  despite the fact that they are not subject to the addicional theoretical uncertainty relating form factors to skin thicknesses.  

Another study by Ref.~\cite{Reinhard2022} which also considered dipole polarizabilities concluded that no single interaction could satisfy both PREX and CREX interactions to 68\% confidence, agreeing with this paper and Ref.~\cite{Zhang2022}, but did not attempt to infer a 'best-fit' interaction.

An analogous analysis for $K_N$ is shown in Figure~\ref{fig:capbsvl}.  The confidence regions for both approaches are compatible with that of PNM $\chi$EFT, although the results using PREX+CREX measurements have nearly the same centroid.

\section{Other Nuclear Methods}
\subsection{Correlations from Nuclear Dipole Polarizabilities}
Nuclear measurements involving the giant dipole resonance enable additional constraints on the symmetry energy and can shed light on the somewhat disparate results of recent skin thickness measurements. 
Measured dipole polarizabilities for the neutron-rich nuclides $^{48}$Ca, $^{120}$Sn and $^{248}$Pb are given in Table~\ref{tab:dipole}.
Support for the experimental values comes from theoretical calculations of $\alpha_D$ by Ref.~\cite{Roca-Maza2015}, who found the linear relations:
\begin{eqnarray}
\alpha_D^{48}&\simeq&(0.10\pm0.01)\alpha_D^{208}+0.36\pm0.07{\rm~MeV~fm^3},\cr
\alpha_D^{208}&\simeq&(2.2\pm0.1)\alpha_D^{120}+0.1\pm0.5{\rm~fm}^3,
\label{eq:alpha48}\end{eqnarray}
with correlation coefficients of 0.82 and 0.96, respectively.  These relations predict the values $\alpha_D^{48}=2.32\pm0.21$ fm$^3$, slightly larger than the experimental value, from the measured value of $\alpha_D^{208}$, and $\alpha_D^{208}\simeq19.0\pm1.2$ fm$^3$, slightly smaller than the experimental value, from the measured value of $\alpha_D^{120}$, but both within one standard deviation. 
 
\begin{table}[]
\center\caption{Dipole polarizabilities with $1\sigma$ uncertainties.
$^\dagger$$\alpha_D$ values corrected as per Ref.~\cite{Roca-Maza2013}.
\label{tab:dipole}}
\begin{tabularx}{\textwidth}{CCC}
\toprule
\textbf{Nuclide}&\textbf{Reference}&\textbf{\boldmath$\alpha_D$ (fm$^3$)}\\
\midrule
$^{208}$Pb&\cite{Tamii2011}&$19.6\pm0.6^\dagger$\\
$^{120}$Sn&\cite{Hashimoto2015}&$8.59\pm0.37^\dagger$\\
$^{48}$Ca&\cite{Birkhan2017}&$2.07\pm0.22$\\
\bottomrule
\end{tabularx}
 \end{table}
 
 %\vspace*{.25cm}
 Ref.~\cite{Trippa2008} determined that the central energy of the giant dipole resonance of $^{208}$Pb and the symmetry energy has its greatest correlation at the density $n_1=0.1$ fm$^{-3}$, and from the measured value of $\alpha_D^{208}$ thereby deduced the symmetry energy at that density to be $S(u_1)=24.1\pm0.8$ MeV, which is in agreement with $S(u_1)=25.7\pm1.4$ MeV deduced from the masses of closed shell nuclei~\cite{Brown2000}.  Similarly, in a study using 62 non-relativistic and relativistic interactions fitted to ground-state properties of finite nuclei, Ref.~\cite{Zhang2015} deduced that the electric dipole polarizability of $^{208}$Pb has its greatest correlation with the symmetry energy at the density $n_2=0.05$ fm$^{-3}$, and found the symmetry energy at that density to be  $S(u_2)=16.54\pm1.00$ MeV (after correcting the measured dipole polarizability as per Ref.~\cite{Roca-Maza2013}).    Using the expansion Equation (\ref{eq:s}), employing empirical correlations to eliminate $K_{\rm sym}$ and $Q_{\rm sym}$, and including uncertainties in $n_s$, these results imply the two linear relations
\begin{equation}
L=10.9S_V-288\pm16{\rm~MeV}, \qquad L=8.1S_V-207\pm12{\rm~MeV},
\label{eq:diplsv}\end{equation}
respectively.  These are in essential agreement with the correlation derived from nuclear mass measurements.  The dipole polarizability correlations are also  consistent with the relevant confidence regions established from $\chi$EFT as well as from neutron skin thickness measurements.  

Ref.~\cite{Roca-Maza2013} showed the existence of a theoretical correlation between the neutron skin thickness and the electric dipole polarizability, justified by the nuclear droplet model.  For $^{48}$Ca, $^{120}$Sn and $^{208}$Pb, these are~\cite{Roca-Maza2015}
\begin{eqnarray}
\alpha^{48}_DS_V&=&(355\pm44)(r^{48}_{np}/{\rm fm})+12\pm19{\rm~MeV~fm}^3,\cr
\alpha^{120}_DS_V&=&(1234\pm93)\,(r^{120}_{np}/{\rm fm})+115\pm36{\rm~MeV~fm}^3,\cr
\alpha^{208}_DS_V&=&(1922\pm73)\,(r^{208}_{np}/{\rm fm})+301\pm32{\rm~MeV~fm}^3.
\label{eq:alpha-r}\end{eqnarray}
We found similar correlations for $^{208}$Pb using calculations of Ref.~\cite{Reinhard2021} and Ref.~\cite{Piekarewicz2021}, respectively:
\begin{eqnarray}
\alpha^{208}_DS_V&=&2195\,(r^{208}_{np}/{\rm fm})+258\pm16{\rm~MeV~fm}^3,\cr
\alpha^{208}_DS_V&=&2125\,(r^{208}_{np}/{\rm fm})+226\pm12{\rm~MeV~fm}^3.
\label{eq:alpha-r1}\end{eqnarray}

Combining relations in Equation~(\ref{eq:alpha-r}) with experimental values for the dipole polarizabilities  (Table~\ref{tab:dipole}) gives the relation 
\begin{equation}
    r_{np}^{48}\simeq(0.056\pm0.056){\rm~ fm}+(0.572\pm0.097)r_{np}^{208} \end{equation}
which compares favorably with Equation~(\ref{eq:capb}), albeit with larger uncertainties, emphasizing the consistency of polarizability and skin experimental results with theory.
The same result is achieved using Equation~(\ref{eq:alpha-r1}) instead.   Equations~(\ref{eq:alpha-r}) and (\ref{eq:alpha-r1}) also suggest that measurements of $r_{np}$ and $\alpha_D$ for $^{48}$Ca, $^{120}$Sn or $^{208}$Pb provide constraints on $S_V$ independently of $L$.  The mean experimental data for $^{48}$Ca, $^{120}$Sn ($r_{np}^{120}$ values are given in Table~\ref{tab:skinSn}) and $^{208}$Pb yield $S_V=29.3\pm10.4$ MeV, $32.1\pm5.2$ MeV, and $31.6\pm2.6$ MeV, respectively, using Equation~(\ref{eq:alpha-r}).  Equation~(\ref{eq:alpha-r1}) alternatively yields $S_V=31.8\pm2.3$ MeV and $29.5\pm2.1$ MeV for $^{208}$Pb.  Collectively, one finds $S_V=30.9\pm1.3$ MeV, which, once again, is consistent with mass fitting, $\chi$EFT and the mean neutron skin results.
 None of these ranges for $S_V$ are consistent with the PREX measurement by itself. 

\begin{table}[H]
\caption{$^{120}$Sn  neutron skin measurements  with $1\sigma$ uncertainties  
\label{tab:skinSn}}
\begin{tabularx}{\textwidth}{cCC}
\toprule
\textbf{\boldmath$^{120}$Sn}&\textbf{Reference}&\textbf{\boldmath$r_{np}^{120}$ (fm)}\\
\midrule
Elastic $p$ scattering&\cite{Terashima2008}&$0.147\pm0.033$\\
Spin-dipole resonance ($^3$He-t)&\cite{Krasznahorkay1999}&$0.18\pm0.07$\\
$\bar p$ annihilation&\cite{Trzcinska2001}&$0.12\pm0.02$\\
\midrule
$^{120}$Sn experimental weighted mean&&$0.130\pm0.017$\\
\bottomrule
\end{tabularx}
\end{table}

\subsection{Correlations from Heavy Ion Collisions}

Two general problems with extracting symmetry energy parameters from
heavy-ion collisions are: (1)  matter in these collisions is close to symmetric, with the symmetry
energy being perhaps 10\% of the total and therefore difficult to
ascertain; and (2) the matter has excitation energies of 50-several hundred MeV per baryon (corresponding to \mbox{$T\sim20--50$ Mev}). This paper has therefore concentrated on probes connected
with cold finite nuclei where densities are near $n_s$ and a Taylor expansion is appropriate.
Here, some results from analyses of heavy ion experiments are summarized.

It has been proposed~\cite{Xie2021} to use empirical pressures deduced
in cold SNM from hadronic transport model analyses of heavy-ion
collisions~\cite{Fuchs2006,Danielewicz2002}.  Data come from
studying kaon production~\cite{Laue1999,Sturm2001}, which provides
constraints in the density range from \mbox{$1.3n_s$ to $2.2n_s$}, and nuclear
collective
flow~\cite{Partlan1995,Liu2000,Pinkenburg1999,Braun-Munzinger1998},
which provides constraints from \mbox{$2.0n_s$ to $3.7n_s$}.  Pressures from kaon
production (collective flows) have estimated uncertainties of about
$\pm22--25\% (\pm19\%--32\%)$, so such constraints have appreciable
errors even before systematic uncertainties, such as from
extrapolating from moderate excitation energies to zero temperature, are taken into account.
In addition, these estimates are based upon Taylor expansions of the
symmetric matter energy like Equation~(\ref{eq:s2}), but including a
fourth-order (kurtosis $Z$) term.  This introduces additional
uncertainty, not only because of questionable validity of such an expansion at high density, but also because the
expansion is arbitrarily truncated at fourth order. Although interesting correlations betweeen $K_{1/2}$ and $Q_{1/2}$ are deduced from this analysis, application to PNM and therefore extraction of symmetry energy parameters becomes model dependent. 

The symmetry energy might more directly probed in heavy ion collisions
through the $\pi^-/\pi^+$ multiplicity ratio in central
collisions~\cite{Xiao2009,Feng2010,Xie2013} and from isospin diffusion or elliptic flows of
neutrons and protons~\cite{Tsang2004,Liu2007,Famiano2006,Cozma2013}; see Ref.~\cite{Yong2020} for a
summary.  Both approaches, however, have given rather conflicting
results concerning the stiffness of the symmetry energy, with $L$
ranging from 0 MeV to $\sim120$ MeV, indicating a large degree of
model dependence in the analyses~\cite{Tsang2009,Cozma2021,Jhang2021}.

For example, the analysis of Ref.~\cite{Tsang2009}, which assumes that the density-dependent symmetry energy has the simple form
\begin{equation}
S(u)=C_1u^{2/3}+(S_V-C_1)u^\gamma,\label{eq:s3}\end{equation}
where $C_1\simeq12.5$ MeV, predicts a $2\sigma$ band in $S_V-L$ space which is shown as HIC in Figure~\ref{fig:correl4}.   For $S$ in Equation~(\ref{eq:s3}), a line in the $S_V-L$ plane with a slope $\Delta L/\Delta S_V$ implies that $S$ is best determined at density $u_b=\exp(-3\Delta S_V/\Delta L)$.  In addition, any point in this diagram implies a $\gamma$ value
\begin{equation}
\gamma={L-2C_1\over3(S_V-C_1)}.
\label{eq:gamma}\end{equation}
The diagonal sides of the quadrilateral in Figure~\ref{fig:correl4} therefore imply that $0.28\le\gamma\le1.04$ and 
$0.35\le u_b\le0.58$.  Therefore, this experiment best probes subsaturation densities smaller than do mass fits ($u_b\sim0.74$), neutron skin thicknesses ($u_b\sim0.66$) or neutron matter theory ($u_b\sim0.66$).  However, the symmetry parameters are rather poorly constrained (possibly because the extrapolation to saturation density is larger in the HIC case), with the quadrilateral having a much larger extent than the regions determined in Figure~\ref{fig:capbsvl}.  In addition, large portions of the quadrilateral violate the UGC.  Furthermore, another analysis of the same data~\cite{Cozma2021} suggests $L=122\pm57$ MeV, indicating overall difficulties with this method.
%%%%%%%%%%%%%%%%%%%%%%%%%%%%%%%%%%%%%%%%%%
\section{Astrophysical Considerations}
Astrophysical observations of neutron stars can yield estimates of radii, moments of inertia, and tidal deformabilities.  We will consider two of the currently most popular observations: the LIGO/Virgo detection of the neutron star merger GW170817~\cite{Abbott2017,De2018} and NICER observations of the rapidly rotating pulsars PSR J0030+0451~\cite{Miller2019,Riley2019} and PSR J0740+6620~\cite{Miller2021, Riley2021}. It will be seen that the symmetry parameters ranges suggested by nuclear experiment and theory are consistent with inferences from astrophysical observations of neutron stars.

\subsection{Neutron Star Radii}
\begin{figure}[h]
\vspace*{-1cm}
\begin{adjustwidth}{-\extralength}{0cm}
\centering
\hspace*{-0.2cm}\hspace*{-1.5cm}\includegraphics[width=10.5cm,angle=180]{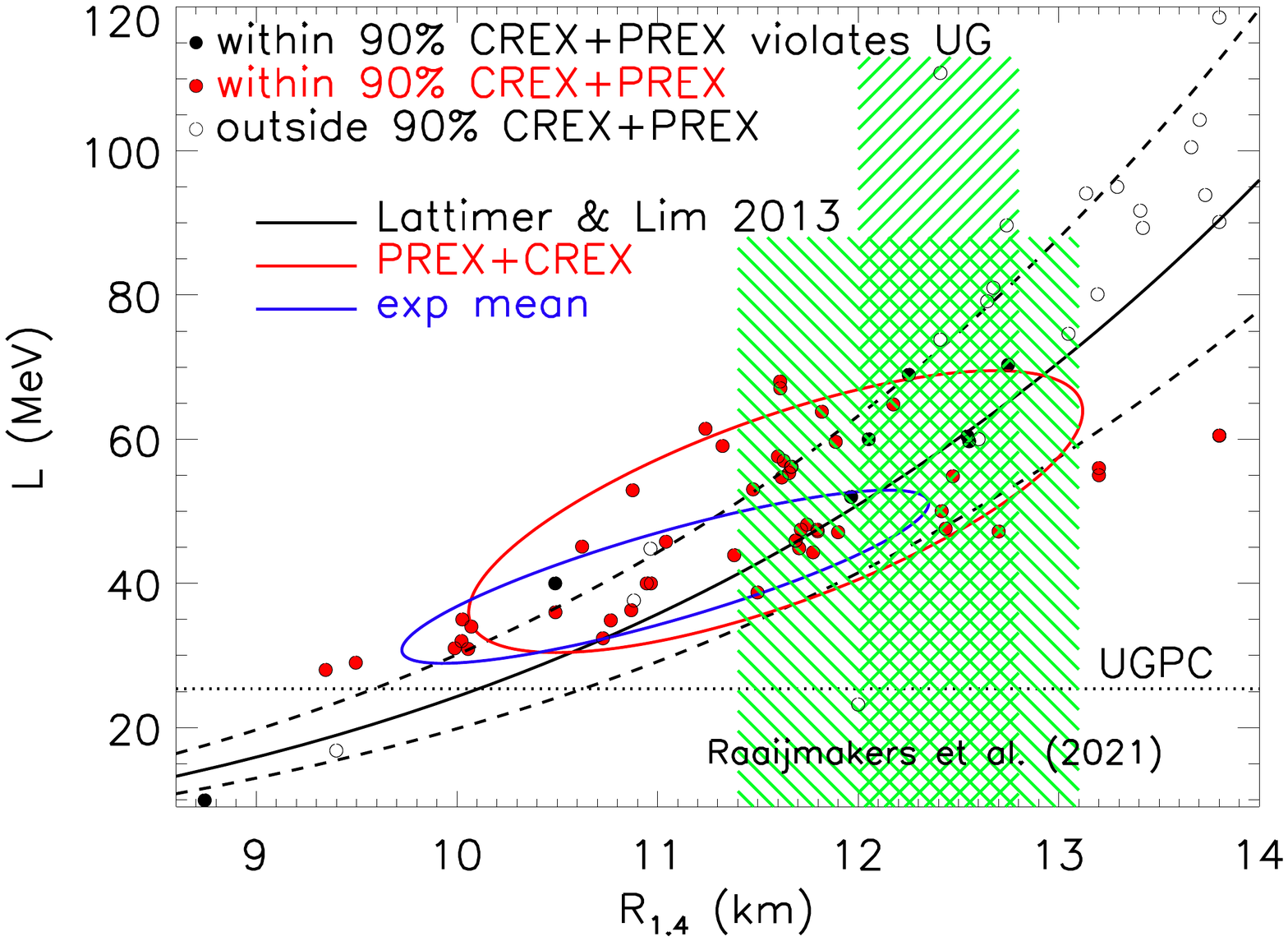}
\hspace*{-2.2cm}\includegraphics[width=10.5cm,angle=180]{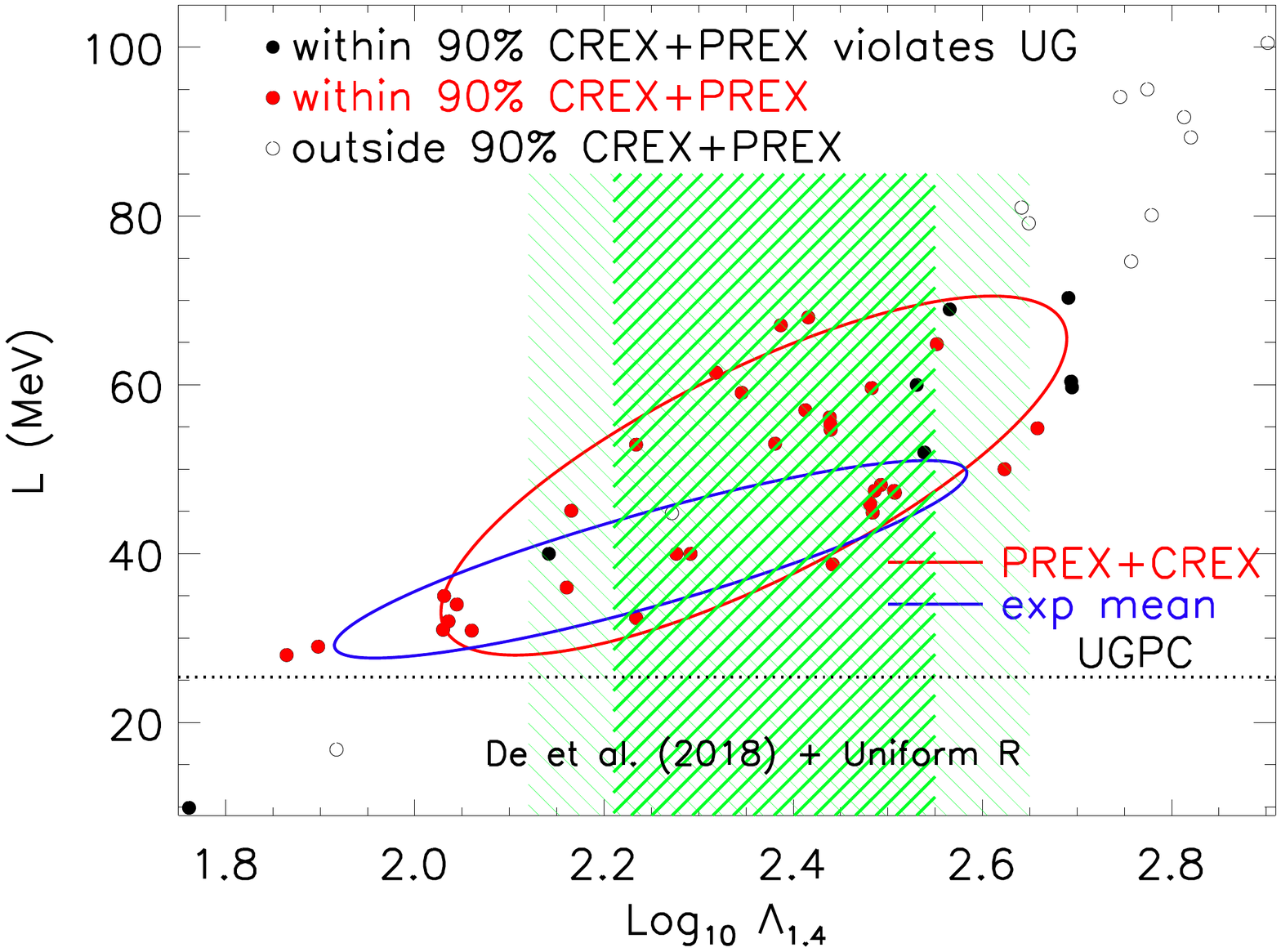}
\end{adjustwidth}
\vspace*{-0.5cm}
\caption{The same as Figure~\ref{fig:capbsvl} except showing $R_{1.4}$ versus $L$ (\textbf{left panel}) and $\Lambda_{1.4}$ versus $L$ (\textbf{right panel}) for subsets of forces from Refs.~\cite{Tagami2022,Adhikari2022,Brown2017,Horowitz2014,Piekarewicz2012}.  Black solid and dashed curves in the left panel show the $R_{1.4}-L$ correlation and standard deviations derived as in the text. The shaded green bands are 68\% and 95\% confidence intervals from a joint analysis of GW170817 and PSR J0030+0451 and PSR J0740+6620 by Ref.~\cite{Raaijmakers2021} (\textbf{left panel}) and from GW170817 Bayesian analyses posteriors~\cite{De2018}, corrected for $\Lambda$ priors chosen so as to reflect uniform $R$ priors (\textbf{right panel}).}
\label{fig:lr14}\end{figure}

For subsets of forces from Refs.~\cite{Tagami2022,Adhikari2022,Brown2017,Horowitz2014,Piekarewicz2012}, the radii of $1.4M_\odot$ neutron stars ($R_{1.4}), L$ and $r_{np}^{48}$ and $r_{np}^{208}$ are available (most of the radius information was kindly provided by B. Tsang).  Although nuclear interaction models, having been fit to nuclei, describe matter with large proton fractions and $n\simle n_s$, one should be wary of making predictions about neutron stars from them.  Nevertheless, as shown in Figure \ref{fig:lr14}, there is a clear correlation between $L$ and $R_{1.4}$.
The correlation between $P_{NSM}(n)$ and $R_{1.4}$ turns out to be greatest at densities in the range $1.5-2n_s$~\cite{Drischler2021a}, low enough that interaction models are still reliable. Note, however, that these models should not be used to constrain neutron star maximum masses, which are most sensitive to $P_{NSM}$ for $n\simge3n_s$~\cite{Drischler2021a}.  

 One can predict a validity range for $R_{1.4}$ based on neutron skin measurements in the same way the $S_V-L$ predictions shown in Figure~\ref{fig:capb} were made. Weighting the predicted $R_{1.4}$ of models satisfying both Unitary Gas constraints by their double Gaussian probabilities in $r_{np}^{48}-r_{np}^{208}$ space, one finds $R_{1.4}=11.6\pm1.0$ km (and $L=49.4\pm13.1$ MeV) to 68\% confidence using parity-violating neutron skin measurements (Figure~\ref{fig:lr14}).    
In comparison, the weighted average of other experiments yields $R_{1.4}=11.0\pm0.9$ km (and $L=40.7\pm7.9$ MeV). 

Bayesian analyses combining measurements of tidal deformability (see Section~\ref{sec:tidal}) from GW170817 and radii from pulse-profile modeling of PSR J0030+0451 and PSR J0740+6620 suggest $R_{1.4}=12.3^{+0.5}_{-0.3}$ km to 68\% confidence~\cite{Raaijmakers2021}. The 68\% and 90\% confidence intervals are shown by green shading in the left panel of Figure~\ref{fig:lr14}.   These astrophysical inferences are in essential agreement with our results using PREX and CREX data, and are about $1\sigma$ larger then results obtained using other experimental neutron skin data. 

It is possible to understand the correlation observed in Figure~\ref{fig:lr14}.   Since NSM is close to PNM around $n=n_s$, there should a high degree of correlation between $P_{NSM}(n_s)$ and $L$, and therefore between $R_{1.4}$ and $L$.  Indeed, a correlation between $R_{1.4}$ and $P_{NSM}$, the neutron star matter pressure, around $n_s$ was empirically established some time ago by Ref.~\cite{LP01} and refined by Ref.~\cite{Lattimer2013},
\begin{equation}
   R_{1.4}\simeq (9.51\pm0.49)(P_{NSM}/{\rm MeV~fm}^{-3})^{1/4}{\rm~km}.
\label{eq:r14p}\end{equation}
It is necessary to convert Equation~(\ref{eq:r14p})
 into one depending directly upon the symmetry parameters by taking into account how NSM differs from PNM.  NSM has a finite proton (electron) fraction determined by energy minimization with respect to $x$ at every density, leading to beta-equilibrium. 
The total energy per baryon of NSM in the quadratic approximation is
\begin{equation}
    E_{NSM}=E_N-4xS(1-x)+{3\over4}\hbar c x(3\pi^2nx)^{1/3},\label{eq:ensm}
\end{equation}
where $E_N$ is the energy of PNM, $S(n)$ is the density-dependent symmetry energy and $x$ is the proton fraction. The last term is the electron contribution. Beta equilibrium requires $\partial E_{NSM}/\partial x=0$, giving
\begin{equation}
    x=\left(4S\over\hbar c\right)^3{(1-2x)^3\over3\pi^2n}.
\label{eq:xsv}
\end{equation}
At $n_s$, one finds
\begin{equation}
P_{NSM}(n_s)=n^2{\partial E_{NSM}\over\partial n}\bigg|_{n_s}={Ln_s\over3}[1-x+2x^2].
\label{eq:pnsm}\end{equation}
At $n_s$, Only the $L$ term in a density expansion of $S$ contributes.  For a given value of $L$, a normal distribution of $S_V$ values based on the PNM $\chi$EFT correlation (Figure~\ref{fig:correl4}), i.e.,
\begin{equation}
S_V=0.139L+(24.63\pm0.25){\rm~ MeV},
\end{equation}
can be used to find a corresponding $x$ distribution from Equation~(\ref{eq:xsv}).  Then, Equation~(\ref{eq:pnsm}) gives the $P_{NSM}(n_s)$ distribution and Equation~(\ref{eq:r14p}) gives the $R_{1.4}$ distribution, which is shown with uncertainties in Figure~\ref{fig:lr14} as black lines.  The agreement, including uncertainties, with model interactions is remarkable, in spite of the fact that Equation~(\ref{eq:r14p}) was established using a far smaller number of interactions.

\subsection{Tidal Deformabilities and Radii\label{sec:tidal}}
For GW170817, the most accurately-measured quantity is the chirp mass~\cite{Abbott2017}
\begin{equation}
    {\cal M}={(m_1m_2)^{3/5}\over(m_1+m_2)^{1/5}}=1.186\pm0.001M_\odot
\end{equation}
where $m_1>m_2$ are the masses of the merging stars.  Less precisely measured is the radius-sensitive binary tidal deformability of the system, 
\begin{equation}
    \tilde\Lambda={16\over13}{(1+12q)\Lambda_1+(12+q)q^4\Lambda_2\over(1+q)^5},
\end{equation}  
where $q=m_2/m_1$ is the mass ratio and $\Lambda_1$ and $\Lambda_2$ are the individual stellar deformabilities. Typical $M(R)$ trajectores in the relevant mass range $1.1M_\odot<M<1.6M_\odot$ for BNS merger components (judging from galactic BNS~\cite{Lattimer2012,Lattimer2021}) have relatively small variations in $R$ for a given EOS. Given the minimum (maximum) neutron star mass is about $1.1M_\odot$ ($2.1M_\odot$), $0.7\simle q\le1$.   Ref.~\cite{Zhao2018} found, using piecewise polytropes, that in the absence of significant phase transitions, the average radius $\bar R$ satisfies $|\bar R-R|<0.5$ km for all EOSs in this mass range, and averaged over all EOSs and masses, $|\bar R-R|$ has a standard deviation of about 0.1 km.   Ref.~\cite{Zhao2018} also determined, to high accuracy, that the dimensionless tidal deformability of individual stars in this mass range obeys the semi-universal (EOS-independent) relation
\begin{equation}
    \Lambda=a\beta^{-6};\qquad  \Lambda_1\simeq q^6\Lambda_2.
        \label{eq:q6}
    \end{equation}
    with $a=0.009\pm0.002$ and $\beta=GM/Rc^2$.  
        It therefore follows that
\begin{equation}
\tilde\Lambda_{\cal M}\simeq{16a\over13}\left({\bar Rc^2\over G{\cal M}}\right)^6q^{8/5}{(12-11q+12q^2)\over(1+q)^{26/5}},
\label{eq:tillam1}\end{equation}
where the notation $\tilde\Lambda_{\cal M}$ reflects that $\tilde\Lambda$ is measured at a well-defined ${\cal M}$.
The $q$-dependence in Equation~(\ref{eq:tillam1}) is weak: $\tilde\Lambda_{\cal M}(q=0.7)/\tilde\Lambda_{\cal M}(q=1.0)=1.029$ and  $(\partial\tilde\Lambda_{\cal M}/\partial q)_{q=1.0}=0$.  Thus, the fact that $q$ is poorly measured has little consequence, and one finds
\begin{equation}
\bar R\simeq(11.3\pm0.3){{\cal M}\over M_\odot}\left({\tilde\Lambda_{\cal M}\over800}\right)^{1/6}{\rm~km}. 
\label{eq:tillam2}\end{equation}
Due to radius correlations, the uncertainty in $\bar R$ is only half the uncertainty of $a$ itself.  Thus, $\tilde\Lambda_{\cal M}$ carries significant radius information.  To the same accuracy, $R_{1.4}\simeq\bar R$.

More directly, one can predict $\Lambda_{1.4}$, the dimensionless tidal deformability of a $1.4M_\odot$ neutron star, from neutron skin measurements using the same procedure as used for $R_{1.4}$.  
 Values of $\Lambda_{1.4}$ for the interactions in the right panel of Figure~\ref{fig:lr14} were provided by B. Tsang.  Weighting each by the double Gaussian defined by neutron skin thickness measurements and uncertainties, it is found, to 68\% confidence,  $\Lambda_{1.4}=228^{+148}_{-90}$ (\mbox{$L=49.3\pm14.0$ MeV}) using the PREX+CREX neutron skin measurements, and  $\Lambda_{1.4}=177^{+117}_{-70}$ (\mbox{$L=39.3\pm7.7$ MeV}) using the average of other skin measurements.  
 
 These ranges are consistent with inferences~\cite{De2018} from LIGO/Virgo observations~\cite{Abbott2017,Abbott2019} of the BNS merger GW170817 (shaded green regions in the right panel of Figure~\ref{fig:lr14}).  For this comparison, it is useful to convert the observed quantity $\tilde\Lambda_{\cal M}$ to $\Lambda_{1.4}$.  Use of Equation~(\ref{eq:q6}) gives
\begin{equation}
    \Lambda_M\simeq\left({2^{1/5}{\cal M}/ M}\right)^6\tilde\Lambda_{\cal M}.
\end{equation}
For GW170817, $\Lambda_{1.4}\simeq0.849\tilde\Lambda_{1.186}$.
Ref.~\cite{De2018} calculated the tidal deformability posteriors of the two merging stars from a grvitational wave analysis assuming $\Lambda_1/\Lambda_2=q^6$ and causality.  They assumed uniform $\Lambda$ and $M$ priors for the two stars.  They found $\Lambda_{1.4}=245^{+312}_{-81}$ to 68\% confidence and $\Lambda_{1.4}=245^{+651}_{-114}$ to 90\% confidence. However, the results are sensitive to the assumed $\Lambda$ priors.  Perhaps it is more realistic to assume uniformity in $\ln\Lambda$ rather than $\Lambda$ because $\Lambda\propto (R/M)^6$ has a large dynamic range.  This is tantamount to a uniform radius prior.  In this case, to 68\% confidence, $\Lambda_{1.4}=234^{+119}_{-72}$, and to 90\% confidence, $\Lambda_{1.4}=234^{+216}_{-101}$. Although the predicted means are nearly the same, assuming uniformity in the $\ln\Lambda$ priors dramatically reduces the upper 68\% and 90\% confidence bounds, and makes them strikingly similar to those inferred from neutron skin measurements.

\section{Conclusions}

Compilations containing several hundred theoretical nuclear interactions of both the non-relativistic (Skyrme) or relativistic (RMF) type, which have been fitted to nuclear masses and/or charge radii, show that significant correlations exist among symmetry and neutron energy parameters $S_V$, $L$, $K_N$ and $Q_N$. The average values and correlation slopes depend on whether the interactions considered are non-relativistic or relativistic: the relativistic interactions in these compilations tend to have larger values of $L$ and $K_N$, for example.   In both cases, the 68\% confidence ellipses for the $S_V-L$, $L-K_N$ and $Q_N-L$ correlations considerably overlap with those predicted by chiral effective field calculations of pure neutron matter, but have different slopes.

Theoretical calculations of neutron skin thicknesses of neutron-rich closed shell nuclei ($^{48}$Ca and $^{208}$Pb) predict strong correlations with $L$ with average uncertainties less than 0.01 fm.  The individual weighted means of experimental measurements of these nuclei can be used to predict comparable ranges of $L$,  0-40 MeV for $^{48}$Ca and 30-60 MeV for $^{208}$Pb, both consistent with the range determined by jointly fitting masses and neutron matter calculations.  Without a strong reason to prefer skin measurements of one nucleus over another, simultaneous reconciliation of the weighted means of skin measurements of both nuclei with those from theoretical calculations of forces in the compilations constrains the symmetry parameters $S_V=30.8\pm1.5$ MeV and $L=40\pm8$ MeV. 

Alternatively, one could choose to only employ parity-violating electron scattering results (PREX and CREX), which have been claimed to have smaller systematic uncertainties than other techniques. The two experiments separately predict incompatible ranges of $L$ ($L=-5\pm40$ MeV and $L=121\pm47$ MeV, respectively), but accepting both measurements to be equally valid suggests $S_V=32\pm2$ MeV and $L=50\pm12$ MeV.  These alternative approaches in treating neutron skin measurements predict closely overlapping ranges that are consistent with ranges predicted from mass fits and neutron matter theory, and therefore indicate that the PREX and CREX experimental results are consistent with those of other techniques. These estimates are compatible with those calculated by Ref.~\cite{Zhang2022}.

Further support for the resulting focus to small ranges of symmetry parameters comes from measurements and theoretical studies of nuclear dipole polarizabilities ($\alpha_D$).  Theory predicts that $\alpha_DS_V$ is linearly correlated to $r_{np}$. Measured values of $\alpha_D$ and $r_{np}$ for $^{48}$Ca, $^{120}$Sn and $^{208}$Pb then imply $S_V=30.9\pm1.3$ MeV in good agreement with our result.  Further proof that existing measurements of $\alpha_D$ and $r_{np}$ are compatible comes from the  linear relation between $r_{np}^{48}$ and $r_{np}^{208}$ resulting from the theoretical $\alpha_DS_V-r_{np}$ correlation; this linear relation is in remarkable agreement with that predicted by theoretical neutron skin calculations of these nuclei using both relativistic and non-relativistic interactions.

The inferred small ranges of symmetry parameters combining mass fitting, neutron matter theory, and skin thickness and dipole polarizability measurements and theory are robust and do not depend significantly on the type of nuclear interaction considered, despite the systematic differences in predictions of individual quantities arising from a single method alone.  Potentially, an improvement of our predictions can be achieved by directly comparing predictions and measurements of the charge and weak form factors of $^{48}$ Ca and $^{208}$Pb, as done in Ref.~\cite{Zhang2022}.  This would remove the additional uncertainty involved in converting form factors to neutron and proton radii.  However, given the compatibility of results from our work with Rev.~\cite{Zhang2022}, this additional uncertainty does not appear to be significant.  In any case, measurements of form factors or neutron skin thicknesses will have to be performed with much greater precision than at present to improve upon  the symmetry parameter constraints from chiral Lagrangian $chi$EFT calculations.

For the first time, we extended predictions from neutron skin measurements to include neutron star radii and tidal deformabilities without employing any information from astrophysical observations.  Our estimated ranges for $R_{1.4}$ and $\Lambda_{1.4}$ are compatible with recent studies using LIGO/Virgo data from the BNS merger GW170817 and NICER X-ray observations of PSR J0030+0451 and PSR J0740+6620. 
%%%%%%%%%%%%%%%%%%%%%%%%%%%%%%%%%%%%%%%%%%

\vspace{6pt}
\funding{This work was initiated through the NSF-funded Physics Frontier Center Network for Neutrinos, Nuclear Astrophysics, and Symmetries (N3AS) and was supported by the U.S.~DOE grant
DE-AC02-87ER40317.}

\dataavailability{Not applicable.}

\acknowledgments{Thanks are due to J. Piekarewicz, W. Nazarewicz and B. Tsang for providing data and, together with T. Zhao, C. J. Horowitz and K. Kumar, helpful discussions. }

\conflictsofinterest{The author declares no conflicts of interest.  The funders had no role in the design of the study, the collection, analyses, or interpretation of data, the writing of the manuscript, or in the decision to publish the results.}

%%%%%%%%%%%%%%%%%%%%%%%%%%%%%%%%%%%%%%%%%%
%% Optional
\abbreviations{Abbreviations}{
The following abbreviations are used in this manuscript:\\

\noindent 
\begin{tabular}{@{}llll}
NS & neutron star & BNS & binary neutron stars\\
EOS & equation of state & PSR & pulsar\\
SNM & symmetric nuclear matter & PNM & pure neutron matter\\
$\chi$EFT & chiral effective field theory & RMF & relativistic mean field\\
UGC & Unitary Gas Conjecture & UGPC & Unitary Gas Pressure Conjecture\\
&&NICER & Neutron Star Interior Composition ExploreR
\end{tabular}}

%%%%%%%%%%%%%%%%%%%%%%%%%%%%%%%%%%%%%%%%%%
%% Optional
%\appendixtitles{no} % Leave argument "no" if all appendix headings stay EMPTY (then no dot is printed after "Appendix A"). If the appendix sections contain a heading then change the argument to "yes".
%\appendixstart

%%%%%%%%%%%%%%%%%%%%%%%%%%%%%%%%%%%%%%%%%%
\begin{adjustwidth}{-\extralength}{0cm}

\reftitle{References}


%\bibliography{your_external_BibTeX_file}

%=====================================
% References, variant B: internal bibliography
%=====================================

%%%%%%%%%%%%%%%%%%%%%%%%%%%%%%%%%%%%%%%%%%
%% for journal Sci
%\reviewreports{\\
%Reviewer 1 comments and authors’ response\\
%Reviewer 2 comments and authors’ response\\
%Reviewer 3 comments and authors’ response
%}
%%%%%%%%%%%%%%%%%%%%%%%%%%%%%%%%%%%%%%%%%%
\end{adjustwidth}
\end{document}